  \def\versionno{ DIA -- version 3 -- by jf -- 20.9.95 }

\def\dl {\bf}

\catcode`\@=11

\newif\if@fewtab\@fewtabtrue
{\count255=\time\divide\count255 by 60
\xdef\hourmin{\number\count255}
\multiply\count255 by-60\advance\count255 by\time
\xdef\hourmin{\hourmin:\ifnum\count255<10 0\fi\the\count255}}
\def\ps@draft{\let\@mkboth\@gobbletwo
    \def\@oddhead{}
    \def\@oddfoot{\hbox to 7 cm{\tiny \versionno
       \hfil}\hskip -7cm\hfil\rm\thepage \hfil}
    \def\@evenhead{}\let\@evenfoot\@oddfoot}

\def\draftcite#1{\ifnum\draftcontrol=1#1\else{}\fi}
\def\@lbibitem[#1]#2{\item{}\hskip -3cm \hbox to 2cm
{\hfil$\scriptstyle\draftcite{#2}$}\hskip
1cm[\@biblabel{#1}]\if@filesw
     {\def\protect##1{\string ##1\space}\immediate
      \write\@auxout{\string\bibcite{#2}{#1}}}\fi\ignorespaces}

\def\@bibitem#1{\item\hskip -3cm \hbox to 2cm
{\hfil {\footnotesize\draftcite{#1}}}\hskip 1cm
\if@filesw \immediate\write\@auxout
       {\string\bibcite{#1}{\the\value{\@listctr}}}\fi\ignorespaces}

\def\citen#1{\if@filesw \immediate\write \@auxout {\string\citation{#1}}\fi%
\@tempcntb\m@ne \let\@h@ld\relax \def\@citea{}%
\@for \@citeb:=#1\do {\@ifundefined {b@\@citeb}%
    {\@h@ld\@citea\@tempcntb\m@ne{\bf ?}%
    \@warning {Citation `\@citeb ' on page \thepage \space undefined}}%
    {\@tempcnta\@tempcntb \advance\@tempcnta\@ne
    \setbox\z@\hbox\bgroup\ifcat0\csname b@\@citeb \endcsname \relax
    \egroup \@tempcntb\number\csname b@\@citeb \endcsname \relax
    \else \egroup \@tempcntb\m@ne \fi \ifnum\@tempcnta=\@tempcntb
    \ifx\@h@ld\relax \edef \@h@ld{\@citea\csname b@\@citeb\endcsname}%
    \else \edef\@h@ld{\hbox{--}\penalty\@highpenalty
    \csname b@\@citeb\endcsname}\fi
    \else \@h@ld\@citea\csname b@\@citeb \endcsname \let\@h@ld\relax \fi}%
\def\@citea{,\penalty\@highpenalty\hskip.13em plus.13em minus.13em}}\@h@ld}
\def\@citex[#1]#2{\@cite{\citen{#2}}{#1}}%
\def\@cite#1#2{\leavevmode\unskip\ifnum\lastpenalty=\z@\penalty\@highpenalty\fi%
  \ [{\multiply\@highpenalty 3 #1%
  \if@tempswa,\penalty\@highpenalty\ #2\fi}]}   %
\makeatother 

\newcommand\Ac[2]  {A^{#1,#2}_{}}

\def\aff           {affine Lie algebra}
\def\alg           {algebra}
\def\alphab        {{\bar\alpha}}

\newcommand\als[1] {\mbox{$\alpha^{(#1)}$}}

\def\AO            {\brev{A}}
\def\aO            {\brev{a}}
\newcommand\AOc[2] {\brev A^{[#1],[#2]}_{}}
\newcommand\ASect[2]{\appendix\sect{#1}\label{s.#2}}

\def\be            {\begin{equation}}
\def\bearl         {\begin{array}{l}}
\def\bearll        {\begin{array}{ll}}
\def\bearlll       {\begin{array}{lll}}

\def\bll           {\mbox{$b_{\Lambda,\Lambda'}$}}
\def\bllch         {\mbox{$b^{(\ch)}_{\Lambda,\Lambda'}$}}

\def\bllo          {\mbox{$b^{[\omega]}_{\Lambda,\Lambda'}$}}
\def\bllostau      {\mbox{$b^{[\omega]}_{\Lambda,\Lambda'}(-\Frac1\tau)$}}

\def\bmmotau       {\mbox{$b^{[\omega]}_{\mu,\mu'}(\tau)$}}

\def\brev          {\breve}

\def\calS          {{\Sigma}}
\def\calw          {\mbox{${\cal W}_{\g,\h}$}}
\def\calwo         {\mbox{${\cal W}_{\g/\h}$}}
\def\calwoo        {\mbox{${\cal W}_{\g,\h}^{(0)}$}}
\def\CC            {{\cal C}}

\def\cc            {\Dot c{.49}{.31}}
\def\Ccs           {\Dot C{.51}{.41}}
\def\ccs           {\Dot c{.34}{.26}}

\def\Cdot          {\!\cdot\!}
\def\cft           {conformal field theory}
\def\cfts          {conformal field theories}
\def\ch            {\Psi}
\def\che           {\ch_1}
\def\chd           {\ch_3}
\def\chii          {\raisebox{.15em}{$\chi$}}
\def\Chii          {{\cal X}}
\def\chil          {\chii_\Lambda^{[\omega]}}

\def\chilastau     {\chii_\lambda^{[\omega]}(-\Frac1\tau)}
\def\chilastauP    {\chii_\lambdaP^{[\omP]}(-\Frac1\tau)}

\def\chillch       {\Chii_\ilab^{(\ch)}}

\def\chilP         {\chii_\LambdaP^{[\omP]}}

\def\chimmcht      {\Chii_\jlab^{(\cht)}}
\def\chimtau       {\chii_\mu^{[\omega]}(\tau)}
\def\chimtauP      {\chii_{\mu'}^{[\omP]}(\tau)}

\def\cht           {{\tilde\Psi}}
\def\chz           {\ch_2}

\def\complex       {{\dl C}}

\def\CP            {C'}
\def\CS            {{\cal S}}
\def\csa           {Cartan subalgebra}
\def\CSO           {\brev{\cal S}}
\def\CT            {{\cal T}}
\def\cvira         {coset Virasoro algebra}

\newcommand\Dot[3]
{{#1}\hspace{-#3em}\raisebox{#2em}{$\scriptscriptstyle\bullet$}}

\def\dsum          {\displaystyle\sum}

\def\dyd           {Dynkin diagram}
\def\ee            {\end{equation}}
\def\eE            {{\rm e}}
\newcommand\EE[2]  {E^{#1}_{#2}}
\def\eear          {\end{array}}
\def\eins          {{(1)}}
\def\el            {\ell}

\def\elp           {m}
\def\elta          {\Delta}
\let\emb=\hookrightarrow

\newcommand\erf[1]{(\ref{#1})}
\def\facprop       {factorization property}
\newcommand\fc[3]  {{#1}\hspace{-#3em}\raisebox{#2em}{$\scriptstyle\circ$}}
\newcommand\fcft[3]{{{#1}^{\mskip-#3 mu\raise #2
pt\hbox{$\scriptstyle\circ$}}}}
\def\findim        {finite-dimensional}
\newcommand\fline[1]{\vfill\noindent ------------------\\[1 mm]}
\def\fortria       {for $\tria\iN\{+,\circ,-\}$}
\newcommand\Frac[2]{\mbox{\large$\frac{#1}{#2}$}}

\def\futnot#1      {\ifnum\draftcontrol=1
                   \footnote{~{\sc internal footnote:} #1}\ \fi}
\def\futnote#1     {\footnote{~#1}\ }
\def\g             {{\sf g}}

\def\gb            {\mbox{$\bar\g$}}
\def\gB            {{\bar\g}}
\def\gbP           {\mbox{$\gB'$}}

\def\gbo           {\mbox{$\gB^{}_\circ$}}

\def\gdc           {generalized diagonal coset}

\def\Gid           {\mbox{$G_{\rm id}$}}
\def\GId           {{G_{\rm id}}}
\def\Gidp          {\mbox{$ G_{\rm id}$}}

\def\Gilabel       {G_{\Lambda,\Lambda'}}
\def\Gilabels      {G^\star_{\Lambda,\Lambda'}}
\def\Gjlabel       {G_{M,M'}}
\def\Gjlabels      {G^\star_{M,M'}}
\def\Gklabel       {G_{N,N'}}
\def\gm            {\mbox{$\g^{}_-$}}
\def\gmP           {\mbox{$\g'_-$}}
\def\go            {\mbox{$\g^{}_\circ$}}
\def\goP           {\mbox{$\g'_\circ$}}
\def\gO            {\mbox{$\brev{\g}$}}

\def\Gomilabelz    {G_{\omilablz}}

\def\gP            {\mbox{$\g'$}}
\def\gp            {\mbox{$\g^{}_+$}}
\def\gPO           {\brev\g'}
\def\gPP           {{\g'}}
\def\gpP           {\mbox{$\g'_+$}}
\def\Gstab         {{G_{\Lambda,\Lambda'}}}
\def\Gstabs        {{G^\star_{\Lambda,\Lambda'}}}

\def\gv            {\mbox{$g_{}^\Vee$}}

\def\gz            {\mbox{$\g^{}_{\tria}$}}
\def\gzP           {\mbox{$\g'_{\tria}$}}
\def\h             {{\sf h}}
\def\half          {\mbox{$\frac12\,$}}
\def\hb            {\mbox{$\bar{\sf h}$}}
\def\Hb            {\bar H}

\def\HbP           {\bar H'}
\newcommand\HH[2]{H^{#1}_{#2}}

\def\hitO          {\brev{\it h}}
\def\hl            {\mbox{${\cal H}_\Lambda$}}

\def\hlo           {\mbox{${\cal H}_{\om^\star\Lambda}$}}

\def\hlaP          {\mbox{${\cal H}_{\el;\lambdaP}$}}
\def\hll           {\mbox{${\cal H}_{\Lambda,\Lambda'}$}}
\def\hllch         {\mbox{${\cal H}^{(\ch)}_{\Lambda,\Lambda'}$}}
\def\hlolo         {\mbox{${\cal H}_{\omt\Lambda,\omtP\Lambda'}$}}
\def\hlP           {\mbox{${\cal H}_{\el;\LambdaP}$}}
\def\hLP           {\mbox{${\cal H}_{\LambdaP}$}}
\def\hlPo          {\mbox{${\cal H}_{\el;\omdP\LambdaP}$}}
\def\hlPop         {\mbox{${\cal H}_{\elp;\omdP\LambdaP}$}}
\def\hlPt          {\mbox{${\cal H}_{\taudd\el;\omdP\LambdaP}$}}
\def\hO            {\brev{\h}}

\newcommand\hsp[1] {\mbox{\hspace{#1 em}}}
\def\hw            {highest weight}
\def\hwm           {highest weight module}

\def\hwv           {highest weight vector}
\def\hy            {$\mbox{-\hspace{-.66 mm}-}$}
\def\id            {\sl id}
\def\ii            {{\rm i}}
\def\ihwm          {irreducible highest weight module}
\def\ilab          {{\Lambda,\Lambda'}}
\def\ilabel        {{(\Lambda;\Lambda')}}
\def\ilabelO       {{(\LambdaO;\LambdaO')}}
\def\ilabl         {{[\Lambda;\Lambda']}}

\def\iN            {\!\in\!}

\def\irmod         {irreducible module}

\def\J             {J}

\def\JO            {{\brev J}}
\def\jlab          {{M,M'}}
\def\jlabel        {{(M;M')}}
\def\jlabelO       {{(\brev M;\brev M')}}
\def\jlabl         {{[M;M']}}

\def\klabel        {(N;N')}
\def\klabl         {[N;N']}
\def\kma           {Kac\hy Moo\-dy algebra}

\newcommand\la[1]  {\mbox{$\Lambda_{(#1)}$}}
\newcommand\lab[1] {\mbox{$\Lambdab_{(#1)}$}}
\newcommand\lAb[2] {\mbox{$\Lambdab_{(#1)_{#2}}$}}
\long\def\labl#1   {\label{#1}\ee \ifnum\draftcontrol=1
                   \mbox{ }\\[-12 mm]\query{#1}\\[5 mm] \fi}

\newcommand\lAbP[2]{\mbox{$\Lambdab'_{(#1)_{#2}}$}}
\newcommand\lambdab{\bar\lambda}
\newcommand\Lambdab{\bar\Lambda}

\def\LambdaO       {{\brev\Lambda}}

\def\lambdaP       {{\lambda'}}
\def\LambdaP       {{\Lambda'}}
\def\Lc            {\Dot L{.74}{.48}\hsp{.20}}
\def\Lcs           {\Dot L{.51}{.39}\hsp{.1}}

\def\lie           {Lie algebra}

\def\llb           {\mbox{\large[}}
\def\lLb           {\mbox{\large(}}
\def\Llb           {\mbox{\Large[}}
\def\LLb           {\mbox{\Large(}}

\def\lP            {{\Lambda'}}
\def\LP            {L'}
\def\lrb           {\mbox{\large]}}
\def\lRb           {\mbox{\large)}}
\def\Lrb           {\mbox{\Large]}}
\def\LRb           {\mbox{\Large)}}

\def\mod           {\;{\rm mod}\,}
\def\modinv        {modular invarian}

\def\muP           {\mu'}

\def\mydollar      {$^\pounds$}
\def\Mydollar      {$^{1,\pounds}$}

\def\NP            {\mbox{$N'$}}

\def\ocha          {twining character}
\def\olie          {orbit Lie algebra}
\def\om            {\omega}
\let\omchar=\ocha
\newcommand\omd[1] {{\dot\omega #1}}
\def\omD           {\dot\omega}
\newcommand\omdd[2] {{\dot\omega^{#1} #2}}

\def\omdi          {{\dot\omega i}}
\def\omdj          {{\dot\omega j}}

\def\omdo          {{\dot\omega 0}}

\newcommand\omdP[1]{{\omP{}^\star#1}}
\def\omDP          {\dot\omega'}
\def\omdr          {\omega_3}
\def\omdrD         {\omega_3^\star}
\def\omdrDP        {{\omega_3'}_{}^\star}
\def\ome           {\omega_1}
\def\omh           {\omega_\h}

\def\omilablz      {\omzD\Lambda,\omzDP\Lambda'}

\def\omP           {{\omega'}}
\def\omT           {\mbox{$\omega^\star$}}
\def\oMT           {{\omega^\star}}
\newcommand\omt[1] {{\omega^\star #1}}

\def\omTP          {{\omega'}^\star}
\newcommand\omtP[1]{{{\omega'}^\star #1}}

\def\omtLa         {{\omt\Lambda}}
\def\omz           {\omega_2}
\def\omzD          {\omega_2^\star}
\def\omzDP         {{\omega_2'}_{}^\star}
\def\one           {\mbox{\small $1\!\!$}1}

\def\onehalf       {\mbox{$\frac12$}}

\def\onetoNm       {1,2,...\,,N-1}

\def\otor          {0,1,...\,,r}

\def\otoNm         {0,1,...\,,N-1}

\def\pp            {{(p)}}

\def\qc            {q_{}^{\Lcs_0-\Ccs/24}}

\def\QO            {\brev Q}
\def\qq            {{(p')}}

\long\def\query#1{\hskip 0pt{\vadjust{\everypar={}\small\vtop to 0pt{\hbox{}%
     \vskip -13pt\rlap{\hbox to 50.0pc{\hfil{\vtop{\hsize=8pc\tolerance=6000%
     \hfuzz=.5pc\rightskip=0pt plus 3em\noindent#1}}}}\vss}}}}%

\def\rep           {representation}
\def\Rep           {Representation}
\def\resp          {respectively}
\newcommand\restr[1] {|\raisebox{-.5em}{$#1$}}

\def\rhs           {right hand side}
\def\rll           {\mbox{$R_{\Lambda,\Lambda'}$}}
\def\rlolo         {\mbox{$R_{\omt\Lambda,\omtP\Lambda'}$}}
\def\rO            {{\brev r}}

\newcommand\Scc[4] {{\cal S}^{}_{([#1;#2],\ch),([#3;#4],\cht)}}

\newcommand\sect[1] {\section{#1}\setcounter{equation}{0}}
\newcommand\Sect[2]{\sect{#1}\label{s.#2} \ifnum\draftcontrol=1
\query{s.#2}\fi}
\newcommand\Sf[4]  {\fc{\cal S}{.71}{.45}_{(#1;#2),(#3;#4)}}
\def\sgP           {\mbox{\small$\g'$}}

\newcommand\Skl[4] {{\cal S}^{}_{([#1;#2],\ch_k),([#3;#4],\ch_l)}}

\def\slz           {\mbox{SL(2,\zett)}}
\newcommand\Sm[2]  {S^{}_{#1,#2}}

\def\SNb           {{\cal S}^{[\omega^n]}}
\newcommand\Snb[4] {{\cal S}^{[\omega^n]}_{(#1;#2),(#3;#4)}}
\newcommand\Sno[4] {{\cal S}^{[\omega^0]}_{(#1;#2),(#3;#4)}}

\def\sO            {{\brev s}}
\newcommand\So[2]  {S^{[\omega]}_{#1,#2}}
\def\soa           {strictly outer automorphism}
\def\SOb           {\mbox{${\cal S}^{[\omega]}$}}
\newcommand\Sob[4] {{\cal S}^{[\omega]}_{(#1;#2),(#3;#4)}}
\newcommand\Soo[4] {{\cal S}^{}_{(#1;#2),(#3;#4)}}
\newcommand\SoP[2] {S^{[\omP]}_{#1,#2}}

\def\smat          {$S$-matrix}
\newcommand\SP[2]  {S'_{#1,#2}}

\def\sumGid        {\dsum_{\om\in G_{\rm id}}}

\def\sumGilabel    {\dsum_{\om\in G_{\Lambda,\Lambda'}}}
\def\sumGilabels   {\dsum_{\ch\in G^\star_{\Lambda,\Lambda'}}}

\newcommand\sumpe[1]{\sum_{#1}}

\newcommand\sumqe[1]{\sum_{#1}}

\def\tauD          {\Dot\tau{.49}{.46}_\omega}
\newcommand\taud[1]{\Dot\tau{.49}{.46}_\omega #1}
\newcommand\taudd[1]{\Dot\tau{.34}{.39}_\omega #1}
\def\tauc          {\mbox{$\Dot\tau{.49}{.46}_\omega$}}
\def\tauo          {{\tau_\omega}}
\def\tauoP         {{\tau_\omP}}
\def\tbf           {twining branching function}
\let\tcha=\ocha

\def\tildegv       {\mbox{$\tilde g_{}^\Vee$}}
\def\trhl          {{\rm tr}^{}_{{\cal H}_{\Lambda}}}
\def\trhll         {{\rm tr}^{}_{{\cal H}_{\Lambda,\Lambda'}}}
\def\trhllM        {{\rm tr}^{}_{{\cal H}_{\Lambda,M'}}}
\def\trhllch       {{\rm tr}^{}_{{\cal H}^{(\ch)}_{\Lambda,\Lambda'}}}
\def\trhlP         {{\rm tr}^{}_{{\cal H}_{\el;\Lambda'}}}
\def\trhlPo        {{\rm tr}^{}_{{\cal H}_{\el;\Lambda'(\el) }}}

\def\trhPM         {{\rm tr}^{}_{{\cal H}_{M'}}}
\def\tria          {\#}

\def\twodim        {two-di\-men\-si\-o\-nal}
\def\U             {{\sf U}}

\def\ug            {\mbox{${\sf U}(\g)$}}
\def\ugm           {\mbox{${\sf U}(\g_-)$}}
\def\ugtw          {\mbox{$\tilde{\sf U}(\g)$}}

\def\Vee           {{\scriptscriptstyle\vee}}
\newcommand\version[1] {\ifnum\draftcontrol=1 \typeout{}\typeout{#1}\typeout{}
                   \vskip3mm \centerline{\fbox{\tt DRAFT -- #1 -- \today}}
                   \vskip3mm \fi}

\def\vira          {Virasoro algebra}

\def\vlP           {\mbox{$v^{}_{\el;\LambdaP}$}}
\def\vlPt          {\mbox{$v^{}_{\taudd\el;\omdP\LambdaP}$}}
\def\vlaP          {\mbox{$v^{}_{\el;\lambdaP}$}}

\def\wrt           {with respect to }
\def\wrtt          {with respect to the }
\def\WZW           {Wess\hy Zumino\hy Witten}
\def\wzwm          {WZW model}

\def\wzwts         {WZW theories}
\def\Zbf           {{\dl Z}}
\def\zet           {{\dl Z}}

\def\zett          {\mbox{\small {\dl Z}}}
\def\zwei          {{(2)}}

\global\def\draftcontrol{0}
\catcode`\@=12

\documentstyle[12pt,epsf]{article}

\begin{document}

\begin{flushright}  {~} \\[-23 mm] {\sf hep-th/9509105} \\
{\sf NIKHEF 95-052}\\{\sf DESY 95-173} \\[1 mm]{\sf September 1995}
\end{flushright} \vskip 2mm

\begin{center} \vskip 12mm

{\Large{\bf THE RESOLUTION OF FIELD IDENTIFICATION}} \vskip 0.3cm
{\Large{\bf FIXED POINTS IN DIAGONAL COSET THEORIES}}
\vskip 15mm

{\large J\"urgen Fuchs, \Mydollar \ \,
 Bert Schellekens, $^2$\, \ Christoph Schweigert $^2$}
\\[5mm] {$^1$ \small DESY, Notkestra\ss e 85}
\\      \ {\small D -- 22603~~Hamburg}
\\[5mm] {$^2$ \small NIKHEF-H\,/\,FOM, Kruislaan 409}
\\      \ {\small NL -- 1098 SJ~~Amsterdam}
\end{center}
\vskip 16mm
\begin{quote}{\bf Abstract}.\\
The fixed point resolution problem is solved for diagonal coset theories.
The primary fields into which the fixed points are resolved are described
by submodules of the branching spaces, obtained as eigenspaces of the
automorphisms that implement field identification. To compute the
characters and the modular $S$-matrix we use `orbit Lie algebras' and
`twining characters', which were introduced in a previous paper \cite{fusS3}.
The characters of the primary fields are expressed in terms  branching
functions of twining characters. This allows us to express the modular
$S$-matrix through the $S$-matrices of the orbit Lie algebras associated
to the identification group. Our results can be
extended to the larger class of `generalized diagonal cosets'.
 \end{quote} 
\vfill {}\fline{} {\small {\mydollar}~~Heisenberg fellow}
\newpage
\sect{Introduction}

The coset construction has been proposed already a long time ago as an
important tool to construct \twodim\ \cfts. Surprisingly, however, several
crucial problems concerning the consistency of this construction have not
been solved so far. In this paper we present an answer to one of the key
questions, the problem of field identification and  of the resolution
of field identification fixed points, for an important class of coset
theories, namely those based on diagonal subalgebras. Our results
show in particular that for these coset theories the complete information
needed in the coset construction is already encoded in the underlying \wzwts.
Our methods and results can in fact be easily extended to the larger class
of `generalized diagonal' embeddings.

The main idea of the coset construction is the following. One starts with a
\findim\ \lie\ \gb, which is the direct sum of simple and abelian subalgebras,
i.e.\ a so-called reductive \lie. In addition, one chooses
a subalgebra $\gbP$ of \gb\ which is reductive as well. Using the loop space
construction and incorporating central extensions, one obtains the
corresponding untwisted \kma s \g\ and \gP,
along with a natural embedding of \gP\ into \g. Also,
performing the Sugawara construction, one obtains Virasoro generators $L_n$ for
\g\ and $L_n'$ for \gP, with central charges $c$ and $c'$, \resp.

The crucial observation \cite{goko} underlying the coset construction
is that the generators $\Lc_n:= L_n - L'_n$ satisfy again the commutation rules
of the Virasoro algebra, this time with central charge $\cc := c - c'$; they
are
referred to as the generators as the coset Virasoro algebra. In order
to be able to associate a \cft\ to this Virasoro algebra one would like to find
the representation spaces on which this algebra acts. The hope is, of course,
to express these spaces in terms of \rep\ spaces of $\g$ and $\g'$, for this
allows us  to use the \rep\ theory of affine \lie s, which is by now a
well-developed tool. In this article we will show that this is in fact
possible; however,
in addition to the well-known concepts of the \rep\ theory of affine \lie s,
such as characters and branching functions, one also has to use aspects of
the \rep\ theory that were developed recently, namely the theory \cite{fusS3}
of twining characters and \olie s.

A hint on what these \rep\ spaces might be is given by the observation that the
generators $\Lc_n$ commute with any element of the subalgebra $\gP$,
\be [\Lc_n, J^a_n] = 0 \qquad\mbox{for all}\qquad J^a_n\in\gP\, . \labl{1i}
This suggests the following ansatz for the \rep\ spaces of the coset Virasoro
algebra: fix non-negative integral level for \g\ and take any unitary \hwm\
$\hl$ of $\g$ at this level. Equation \erf{1i} suggests that
one should regard all vectors of $\hl$ which differ only by the action of some
element of $\gP$ as the same vector for the putative \rep\ space of the coset
\cft. We therefore decompose $\hl$ into modules \hLP\ of $\gP$,
$  \hl=\bigoplus_\LambdaP \hll\otimes\hLP\,,$
and tentatively consider $\hll$ as part of the Hilbert space of the coset \cft.
On the level of characters this corresponds to decomposing the character
$\chii^\g_\Lambda$ of \hl\ into characters $\chii^\gPP_{\lP}$ of $\gP$,
according to
  \be \chii^\g_\Lambda(\tau) = \sum_{\lP}\bll(\tau)\, \chii^{\gPP}_{\lP}(\tau)
  \labl{cc}
and taking the {\em branching functions} \bll\ as the characters of
the coset \cft.

This guess, however, turns out to be too naive. Namely, first, several
branching functions vanish, and second, certain non-vanishing branching
functions are identical, in particular the putative vacuum primary field
seems to be present several times \cite{gepn8,levw,mose4}. The main reason
for a branching
function to vanish are conjugacy class selection rules, arising from the
embedding $\gbP\hookrightarrow\gb$ of \findim\ \lie s.

A convenient formalism to implement these two observations is provided by
simple currents (for a review see \cite{scya6}). It can be shown \cite{scya5}
that there is
a subgroup $\Gid$ of the group of integer spin simple currents of the
tensor product theory described by $\g \oplus (\gP)^*$, the so-called
{\em identification group}, such that the group theoretical
selection rules can be expressed by the condition that the monodromy charges
$Q_J$ of any allowed branching function with respect to all simple currents
$J$ in the identification group vanish. Moreover,
\smat\ elements only change by phases on simple current orbits,
  \be  S_{J\star\Lambda,M}=\eE^{2\pi\ii\,Q_J(M)}\cdot S_{\Lambda,M} \,,
  \labl{sjs}
where `$\star$' denotes the fusion product (note that since $J$ is a simple
current, the product $J\star\Lambda$ contains just one primary field so
that the notation in \erf{sjs} makes sense).
By standard simple current arguments \cite{scya6}
these relations and the corresponding relations for the $T$-matrix imply that
there is a modular invariant in which only allowed branching functions occur,
  \be  Z(\tau) = \sum_{\scriptstyle [\Lambda;\lP] \atop\scriptstyle  Q=0}
  |\Gstab| \cdot |\!\!\! \sum_{J\in\GId/\Gstab} \!\!\!
  b_{J\star(\Lambda;\lP)}(\tau)  \,|^2 \, . \labl{2i}
This formula is to be read as follows: the branching functions for which the
monodromy charge $Q$ \wrt all elements of \Gid\ vanishes are organized into
orbits by the action of the group \Gid. The
square bracket indicates that the first summation is over a
representative for each such orbit ($\Lambda$ and $\lP$ are integrable highest
weights of \g\ and \gP, \resp, at the relevant levels). Now for any orbit
we can define a stabilizer group $\Gstab$
which is the subgroup of $\Gid$ consisting of those elements that leave
$(\Lambda;\lP)$ invariant. In the complete square finally we have a sum
over the branching functions in the respective orbit.

Note that the stabilizer of the putative vacuum module, i.e.\ the one with
highest weights $(\Lambda=0,\;\lP=0)$, is trivial, which implies that the
vacuum would appear $|\Gid|^2$ times,
due to the  $|\Gid|^2$ terms in the complete square containing the identity.
Therefore we would like to divide the expression \erf{2i} by this factor.
However, as was first pointed out in \cite{levw}, this will inevitably lead
to problems
such as fractional coefficients in the partition functions
as soon as there are orbits for which the stabilizer is non-trivial.
Such orbits are termed {\em fixed points}.

The factor $|\Gstab|$ in front of the complete square in \erf{2i} suggests
that any fixed point should in fact correspond to $|\Gstab|$ many rather than
to
a single primary field; this is called the {\em resolution} of the fixed point.
As far as \rep s of the modular group are concerned,
a solution of the problem of how to resolve fixed points
was proposed in \cite{scya5}; it prescribes a modification of
the elements of the modular matrix $S$ that involve fixed points in terms
of the $S$-matrix of another (putative) \cft, called the {\em fixed point
theory}. In this description
it seems as if some necessary information for the coset \cft\
is missing which cannot be obtained from data of $\g$ and $\gP$ alone: it
is unclear how the $S$-matrix of the fixed point theory can be interpreted
in terms of the underlying \kma s \g\ and \gP.
A closer analysis also reveals \cite{scya5} that the branching functions
have to be changed as well in order
to obtain the characters of the coset \cft; this is known as
{\em character modification}.
 \futnote{The terms `field identification' and `character modification'
are actually misnomers. Not the fields are to be identified, but rather,
several distinct combinations of highest weights must be identified because
they describe one and the same field of the coset theory; and it is not
the characters which get modified, but rather the branching functions have
to be modified in order that they coincide with
the true characters of the coset \cft.}
 In \cite{scya5} it was checked that the prescription in terms of
the fixed point theory yields
consistent results at the level of \rep s of the modular group \slz.

In this paper, we will use the action of certain outer automorphisms on
the branching spaces to implement field identification and fixed point
resolution for diagonal cosets directly on the candidate Hilbert space of
the theory. The procedure of fixed point resolution is then explained by the
idea \cite{mose4,sche3} that
the branching spaces which correspond to fixed points carry {\em reducible}
\rep s of the coset Virasoro algebra.
These reducible \rep s can be split into  $|\Gstab|$ many irreducible \rep s.

The most important new insight of this paper is that this splitting is
governed by natural structures present in the \kma s \g\ and \gP, namely
by the recently introduced \cite{fusS3} \omchar s and \olie s.
This implies in particular that the fixed point theories which describe this
splitting in fact do not constitute an independent input, but play the role of
summarizing various data that are already obtainable from the algebras
\g\ and \gP. Using the results of \cite{fusS3}, we can show that
at the level of \rep s of the modular group our implementation of fixed
point resolution is consistent with the results of \cite{scya5}. In
particular, we can compute both the modifications of the \smat\ and the
characters explicitly.

This paper is organized as follows. In section \ref{s.emb} briefly review
the construction of diagonal coset \cfts\ and \gdc s and then introduce the
automorphisms relevant for field identification, the so-called diagram
automorphisms. Section \ref{s.bs} is devoted to the implementation of
field identification and a closer characterization of the coset chiral
algebra. To this end branching spaces and branching functions are introduced
and described in subsection \ref{s.bs}.1, and in subsection \ref{s.bs}.2 it is
also explained how each field identification implies a conjugacy class
selection rule as well as an invariance property of the \cvira.
Field identification is then described by the use of certain automorphisms of
\g\ which have the special property that they give also rise to automorphisms
of the subalgebra \gP; we call the latter property the \facprop.
In section \ref{s.fp} we implement fixed point resolution directly on the
branching spaces. In particular, we derive explicit formul\ae\ for the
characters
of the primary fields into which the fixed points are resolved, as well as
a formula
for the \smat\ of the coset \cft. In section \ref{s.bex} we illustrate our
formalism by applying it to diagonal cosets describing $\zet_2$ orbifolds of
the free boson compactified on a circle. Finally, we discuss how our results
can
be generalized to more general cosets.

In two appendices we present some additional information.
In appendix \ref{s.afac} we derive some properties of automorphisms in
general cosets that fulfill the \facprop. Our proof provides in particular
additional insight into the uniqueness of modular anomalies of characters
and branching functions.
In appendix \ref{s.scheck} we show that the \smat\ which is obtained
for the coset theory is indeed a unitary, symmetric matrix,
whose square is a conjugation of the primary fields and which, together with
a diagonal unitary $T$-matrix, generates a \rep\ of the group \slz.

Our results show that the fixed points of coset \cfts\ supply a surprisingly
rich structure. It would be interesting to compare our algebraic approach
to the description
of coset \cfts\ in terms of gauged \wzwm s so as to identify
the analogous structures in the geometric approach. A first step towards
the geometric interpretation of character modifications has been
undertaken recently in \cite{hori}.

\Sect{Embeddings and diagram automorphisms}{emb}

\subsection{Diagonal embeddings}

In this section we briefly recall the definition of diagonal cosets.
We fix a simple \lie\ \hb; then we choose \gb\ as the direct sum of two
copies of \hb, $\gb\cong\hb\oplus\hb$, and as a subalgebra $\gbP$ we take
$\gbP\cong\hb$ embedded diagonally into \gb.
As a general convention, quantities referring to the subalgebra \gP\
will be denoted by the same symbol as the corresponding quantities for \g,
but with a prime added. However, in order to keep the notation at a
reasonable level, we will suppress the
prime whenever the context (such as primed indices) already makes it clear
that one is dealing with an object referring to \gbP.

The loop space construction provides an embedding of the corresponding
untwisted
\kma\ $\gP \cong \h$ into $\g \cong \h \oplus \h$; it is given by
  \be  \EE{\pm\alphab}m= \EE{\pm\alphab}{\eins,m}+\EE{\pm\alphab}{\zwei,m} \,,
  \qquad  \HH im= \HH i{\eins,m}+\HH i{\zwei,m} \, . \labl{diaem}
Moreover, one has to introduce two central extensions $K_\eins$ and $K_\zwei$
for the two ideals of \g\ and a central extension $K'$ for \gP, for which the
embedding is
  \be  K'= K_\eins+K_\zwei  \, . \labl{K12}

Diagonal cosets have been studied extensively in the literature.
Many interesting \cfts\ can be realized in terms of diagonal cosets:
the Virasoro minimal series, the $N=1$ superconformal minimal series
\cite{goko,goko2}, and the rational $\zet_2$ orbifolds of $c=1$ theories.
It was also in the framework of diagonal cosets that extensions of the
chiral algebra have been discussed, leading to realizations of $\cal W$
algebras in terms of affine \lie s (see \cite{Bosc} and references cited
there).
Modular invariants for diagonal cosets have been studied in \cite{chra2,gawa}.

To study the implications of this situation for the \rep\ theory,
let us assume that $V$ is a unitary highest weight module of \g\
on which the central elements $K_\eins$ and $K_\zwei$ act as non-negative
integer multiples $k_\eins, k_\zwei$ of the identity. To obtain also an action
of the Virasoro algebra on $V$, we define $L_m:=L_{(1),m}+L_{(2),m}$, where
  \be L_{(p),m} := \sum_{a,b=1}^{{\rm dim}\, \bar\h}
  \frac{\kappa_{ab}} {2\,(k_{(p)}^{}+\gv)}
  \sum_{n\in\zet} :\!T^a_n T^b_{m-n}\!: \, \ee
is the Sugawara construction of the Virasoro generators of \h\ at level
$k_{(p)}$, i.e.\ the $T^a_n$ are the generators of \h, $\kappa_{ab}$ is
the Killing form of \hb, and the colons `$\,:\ :\,$' denote a normal
ordering.

Due to the embedding $\gP\hookrightarrow\g$
we can view $V$ as a $\gP$-module, on which the central element $K'$
acts as a multiple $k' \one_V$ of the identity where
  \be k' = k_\eins + k_\zwei \, . \ee
To obtain a \rep\ of the Virasoro algebra for \gP\ as well, we also perform
the Sugawara construction for \gP.\,\,%
 \futnote{These Virasoro operators
act in a well-defined way on $V$, because
for any vector $v\in V$ there are only finitely many positive roots of \gP\
for which $E v$, with $E$ the associated step operator, does not vanish.
(Modules with this property are known as `restricted modules'.) Note, however,
that even though $V$ is a unitary highest weight module over \g, it is
typically not a highest weight module over $\gP$ any more, but
rather a (in general infinite) direct sum of unitary \hwm s.}
By direct computation \cite{goko} (compare also \cite{baha}), one checks that
the difference
  \be  \Lc_m := L_m - \LP_m  \labl{cvir}
of the two Virasoro algebras commutes with all elements of $\gP$ and represents
again a Virasoro algebra, with central charge
  \be  \cc := c - c' \,, \labl{refpoint}
the {\em\cvira}.

Diagonal coset theories have the advantage that various properties of the
embedding can be checked in a straightforward manner.
One property of diagonal cosets that can be seen immediately is that
the triangular decomposition
  \be  \g=\gp\oplus\go\oplus\gm   \labl{gbz}
of $\g$ into the \csa\ $\go$ and the Borel subalgebras $\g_\pm$ spanned
by the step operators of positive and negative roots, and the analogous
decomposition
  \be  \gP=\gpP\oplus\goP\oplus\gmP \labl{gzP}
of \gP\ are compatible in the sense that
  \be  \gzP\subseteq\gz \labl{zz}
\fortria. By construction, we then have in fact
  \be  \gzP=\gz\cap\gP \, ,   \labl{hzP}
\fortria.

The results we present in this paper can be generalized
in several directions. First of all, the algebra $\hb$ we use to construct
need not be simple; rather, it can be the direct sum of several simple
and abelian ${\rm u}(1)$ algebras. Moreover, one can also consider embeddings
of more than one copy of $\hb$ into more than two copies of $\hb$.
It is this class of cosets that we call `\gdc s'. However, to keep the
presentation as simple as possible, we will concentrate in this paper
on diagonal cosets and assume that $\hb$ is simple; the extension of our
results to the case of \gdc s is straightforward.

\subsection{Strictly outer automorphisms}

Let us now introduce the class of automorphisms of the \kma\ \g\ which
we will use to solve the fixed point problem.
A {\em strictly outer\/} automorphism (or {\em diagram\/} automorphism) of
an \aff\ \g\ is the automorphism $\om$ that is induced via
  \be  \om(\EE i\pm)=\EE\omdi\pm\,, \qquad
  \om(\HH i{})=\HH\omdi{}\,, \labl{omD}
by an automorphism of the \dyd\ of \g, i.e.\ by a permutation $\omD$
of the labels $i\in\{\otor\}$ of the nodes which leaves the Cartan matrix $A$
invariant, $\Ac\omdi\omdj=\Ac ij$. Since the number of nodes of the \dyd\ is
finite, any strictly outer automorphism has finite order, which we denote by
$N$. As an example for such an automorphism consider the \dyd\ of $A_n^{(1)}$:
it is a regular $n+1$\,-gon; any rotation with an angle a multiple of
$2\pi/(n+1)$ corresponds to the action of a simple current and is a symmetry
of the \dyd.

By construction, any diagram automorphism
respects the triangular decomposition of \g, i.e.\
  \be  \om(\gz)=\gz  \labl{omz}
\fortria. Conversely, if $\omega$ is some automorphism of an affine \lie\ \g\
which obeys \erf{omz}, then $\omega$ is induced by a symmetry $\omD$ of
the \dyd. Namely, let $E^\alpha$ be some step operator corresponding to a
positive root $\alpha$, so that $[h,E^\alpha]=\alpha(h) E^\alpha$ for all
$h\in\go$. Applying the map $\omega$ to both sides of this equation, we
learn that $[\omega(h),\omega(E^\alpha)]=\alpha(h)\, \omega(E^\alpha)$. Since
$\omega(\go)=\go$, this shows that $\omega(E^\alpha)$ is again a step
operator, corresponding to
a root $\omT(\alpha)$, which obeys $\omT(\alpha)(\omega(h))= \alpha(h)$.

The same argument also shows that, in case the root space corresponding to
the root $\alpha$ has dimension larger than one,
this mapping still provides a mapping of root spaces. Thus the automorphism
$\omega$ induces a mapping $\omT$ of the root system; in fact, $\omT$ is
even an automorphism of the root system. Moreover, \erf{omz} implies that
for $\alpha$ a positive root, $\omega(E^\alpha)$ is in $\g_+$, and hence $\omT$
is an automorphism of the root system which maps positive roots on positive
roots. The latter automorphisms are precisely the symmetries of the
\dyd\ of \g\ (compare e.g.\ \cite{FUch}).

The automorphism $\omT$ of the root (or weight) space of \g\ acts on the
simple \g-roots \als i and the fundamental \g-weights \la i as
  \be  \omT(\als i)=\als\omdi \,, \qquad  \omT(\la i)=\la\omdi \,.
  \labl{omT}
Also, there is a unique extension of $\om$ to the semidirect sum of \g\
with the \vira, namely via
  \be  \om(L_m)= L_m- ( \lab{\omd0}, H_m) +
  \onehalf\,(\lab{\omd0},\lab{\omd0}) \,\delta_{m,0}\, K  \labl{omvir}
and $\om(C)=C$, where $C$ is the central element of the Virasoro algebra.
(For a more detailed discussion of these matters we refer the
reader to \cite{fusS3}.)

In the case of coset theories, we are interested in a situation in which the
strictly outer automorphisms of an untwisted \kma\ \g\ and a subalgebra \gP\
are linked. Namely, consider a strictly outer
automorphism $\om$ of \g\ and its restriction
  \be  \omP:=\om\restr{\sgP}  \labl{omP}
to the subalgebra \gP. Assume that $\omP$ has the non-trivial property
that it maps $\gP$ to itself
 \be  \omP:\quad \gP\to\gP  \,;  \labl1
for reasons that will become apparent soon, we
will in this case say that $\omega$ fulfills the {\em\facprop}.

For diagonal cosets the strictly outer automorphisms that fulfill
the \facprop\ can be described explicitly as follows:\,\futnote{For a
 discussion of the \facprop\ in a more general context,
see appendix \ref{s.afac}.}
fix any \soa\ $\omh$ of \h, then $\om:=\omh\oplus\omh$ is a \soa\ of
$\g=\h\oplus\h$; this automorphism restricts to a \soa\ $\omP$ of $\gP=\h$,
which just coincides with $\omh$.
In order to implement field identification and fixed point resolution,
we will use those \soa s that correspond to simple currents;
we remark that if $\omh$ corresponds
to a simple current of \h, then $\omega$ and $\omP$ correspond to the action of
a simple current of $\g$ \resp\ $\gP$.

Also note that any two
diagram automorphisms $\omega_1$ and $\omega_2$ of \g\ corresponding
to simple currents commute,
  \be  [\omega_1,\omega_2]=0 \,. \labl{ooo}
This holds because the action of $\omega_1$ and $\omega_2$ on all of \g\ is
already fully determined by their action on the subalgebra $\g_+$ spanned by
step operators of positive roots, and hence
by their action on the generators corresponding to simple roots \cite{fusS3};
the latter actions, in turn, just correspond to the actions of the
automorphisms $\omD_1$ and $\omD_2$ of the \dyd\ of \g\ which commute.
Analogous statements apply to \gP\ and hence, as we will see later,
if $\omega_1$ and $\omega_2$ fulfill the \facprop, to the action on the
chiral algebra of the coset theory.

\Sect{Branching spaces and selection rules}{bs}

\subsection{Branching spaces}

{}From the fact \erf{1i} that the \cvira\ and the subalgebra \gP\ commute
and from the decomposition \erf{cc} of characters it is apparent
that the objects of basic interest in the coset construction
are the spaces whose `characters' are the branching functions.
As we will see, these spaces, known as {\em branching spaces}, are in fact
the building blocks of the Hilbert space of the coset theory. However,
we emphasize that, in contrast to some claims in the literature,
in the presence of fixed points they do not provide the irreducible
\rep\ spaces of the theory; rather, the construction of the Hilbert space
also requires some of the tools developed in \cite{fusS3}.
More concretely, branching spaces can be introduced as follows.

Any \ihwm\ \hl\ corresponding to a primary field of \g\ decomposes
\cite[\S 12.10]{KAc3}
into (in general infinitely many) \ihwm s of \gP. We write this decomposition
as
  \be  \hl=\bigoplus_\el \hlP  \,. \labl{hh}
Here we label the \gP-modules in the decomposition by an integer
$\el=1,2,...$ and supply as a (redundant) label of these modules their
highest weight $\lP\equiv\lP(\el)$ \wrt \gP. All highest weights $\Lambda'$
appearing in the decomposition have the same level.

Denote by \vlaP\ the \hwv\ of \hlaP, and consider for any \gP-weight
$\LambdaP$ the complex vector space
  \be  \hll:= {\rm span}_\complex\{\vlaP\mid\lambdaP(\el)=\LambdaP\} \,,
  \labl{hll}
i.e.\ the span of those \hwv s \vlaP\ which have weight $\LambdaP$ \wrt \gP\
(or, equivalently, of those \ihwm s of \gP\ appearing in \erf{hh} whose
\hw\ is $\LambdaP$). The space \hll\ is called the {\em branching space}
that corresponds to the combination $(\Lambda;\lP)$ of highest weights.
Using the branching spaces we can write the decomposition \erf{hh} in the
form
  \be  \hl=\bigoplus_\el \hlP \cong \bigoplus_\LambdaP \hll\otimes\hLP\,,
  \labl{hhh}
where the summation is over all integrable highest weights of \gP\ at the
relevant level.
It must be noted that this definition does {\em not\/} guarantee that a
given branching space is non-empty.
We will see later that typically some branching spaces are empty;
the main reason for this are group theoretical selection rules.

To describe field identification, we need to describe maps on branching
spaces. Our strategy will be to construct these maps from maps on
\rep\ spaces of \g\ and \gP; let us therefore describe in some detail
these maps first.
As was shown in \cite{fusS3}, any \soa\ $\om$ of \g\ induces a map
$\tauo\!\!:\; \hl\to\hlo$.
The action of this map on a highest weight module can be described
as follows: denote by $v_\Lambda$ the highest weight vector
of $\hl$ and by $v_{\omtLa}$ the highest weight vector of
${\cal H}_{\omtLa}$. Any vector of $\hl$ is of the form
$R_\Lambda(x) v_\Lambda$,
where $x$ is an element of $\ugm$, the universal enveloping algebra of
$\g_-$, i.e.\ $x$ is a linear combination of products of lowering operators.
Such a vector is mapped by $\tauo$ to the vector
$R_{\omtLa}(\omega(x)) v_{\omtLa}$ in
${\cal H}_{\omtLa}$. Note that we have here $\omega(x)$ acting on
the highest weight vector of some other module; it can be checked \cite{fusS3}
that this prescription ensures that the map $\tauo$ is
compatible with the null vector structure of the corresponding irreducible
highest weight module, i.e.\ $\tauo$ provides a one-to-one correspondence
between the respective null vectors.

{}From this definition it follows that $\tauo$ `$\omega$-twines' the action
of \g\ in the sense that
  \be  \tauo(R_\Lambda(x)\cdot v)= R_\omtLa(\omega(x))\cdot\tauo(v) \,
  \labl{C'}
for all $x\iN\g$ and all $v\iN\hl$, i.e.\ that the diagram
  \be \begin{array}{rcl} \hl&
  \stackrel{R_\Lambda(x)}{\mbox{---------}\!\!\!\longrightarrow}
  & \hl\\[2 mm]
  {\scriptstyle \tauo}\,\downarrow\ && \ \downarrow\,{\scriptstyle \tauo}
  \\[2 mm] \hlo \!\!\!  & \stackrel{R_\omtLa(\omega(x))}
  {\mbox{---------}\!\!\!\longrightarrow} & \!\hlo \end{array}\ee
commutes. {}From now on we will for simplicity drop the symbol `$R_\cdot$'
which
indicates the \rep\ in which the various generators act, as usual (anyway,
in the notations like $x v$, the vector space of which $v$ is an element
already uniquely determines the \rep\ of \g\ that acts on it).

Let us now turn to those automorphisms which fulfill the \facprop; then
$\tauo$  maps \hwv s \wrt \gP\ in the module $\hl$ to \hwv s
\wrt\ \gP\ in the module ${\cal H}_{\omtLa}$. This holds since the
\facprop\ $\om(\gP)\subseteq\gP$ and \erf{hzP} imply that we also have
$\omP(\gpP)\subseteq\gpP$, so that the property
$\gpP\Cdot v=0$ for some $v\in\hl$ implies that also
$\gpP\Cdot(\tauo(v))=\tauo(\om(\gpP)\Cdot v)=\tauo(\gpP\Cdot v)=0$.
Moreover, if $v_1$ and $v_2$ lie in one and the same \ihwm\ of \gP, then so
do $\tauo(v_1)$ and $\tauo(v_2)$. To see this, we can assume that $v_1$ is a
\hwv\ \wrt\ \gP. Then $v_2=u\Cdot v_1$ for
some $u\iN\U(\gmP)$, and hence
$\tauo(v_2)=\tauo(u\Cdot v_1)=\om(u)\cdot\tauo(v_1)$; because of
$\om(u)\iN\U(\gmP)$ this shows that $\tauo(v_2)$ is an element of the Verma
module with \hwv\ $\tauo(v_1)$. Since the map $\tauo$ respects the null vector
structure of the modules, i.e.\ $\tauo(v)$ is a null vector if and
only if $v$ is a null vector, $\tauo(v_2)$ is in fact an element of the \ihwm\
with \hwv\ $\tauo(v_1)$.

Comparing the decomposition \erf{hh} with the analogous decomposition of
$\tauo(\hl)\equiv\hlo$, it then follows that the map $\tauo$ induces a mapping
$\tauD$, of the same order $N$, of the labels of the \gP-modules in \erf{hh},
 \futnot{The order cannot be smaller: take the orbit of the vacuum of \g.
 Then $\tauo$ maps to vectors in $N$ different \ihwm s of \g. Hence all
 the $N$ labels on an orbit of $\tauD$ live in different \g\ spaces and
 therefore have to be different.}
  \be  \el \mapsto \taud\el  \,, \labl{taud}
such that $\tauo(\hlP)=\hlPt$, i.e.\ the decomposition of $\hlo$ reads
  \be  \hlo\equiv\tauo(\hl)=\bigoplus_\el \tauo(\hlP)
  =\bigoplus_\el \hlPt  \,. \labl{hho}
This implies that the action of the map $\tauo$ on an irreducible \g-module
\hl\ can be described as the combination of two maps: first, a map $\tauc$
acting on \gP-submodules of \hl, which maps any irreducible \gP-submodule
of \hl\ to some \gP-submodule of \hlo\ according to
$\hlP\mapsto\tauc(\hlP)=\hlPt=\tauo(\hlP)$ (or in terms of \hwv s,
$\vlP\mapsto\tauc(\vlP)=\vlPt$),
i.e., a map for which, roughly speaking, only the positions of the \gP-modules
$\hlP$ and $\hlPt$ in the respective \g-modules matter;
and second, a map $\tauoP$ (defined via the automorphism $\omP$ of \gP\
in the same way as $\tauo$ is defined via $\om$), which describes how a
specific vector in \hlP, considered just as a \gP-module,
is mapped to a vector in the \gP-module \hlPt.
It is this observation which is the origin of the name \facprop\ that we
chose to describe the situation \erf1.

\subsection{Branching functions}

Since the generators $\Lc_n$ of the \cvira\ commute with all elements of
the subalgebra
$\gP$, they act in a natural way on the branching spaces. We can therefore
introduce the {\em branching functions} $\bll$ as the traces
  \be \bll(\tau) = \trhll \qc \, . \labl{39}
over the branching spaces; here we have set as usual $q:=\exp(2\pi\ii\tau)$.
{}From equation \erf{hhh} and the relations $L_0 = \Lc_0 + L'_0$ it is also
apparent that the branching functions appear in the decomposition
  \be \chii_\Lambda(\tau) = \sum_{\Lambda'} \bll(\tau)\, \chii_{\Lambda'}(\tau)
  \,  \ee
of (Virasoro specialized
 \futnote{Results similar to the ones derived below hold for the full
characters including Cartan angles, provided that for the \g-characters
one includes only the exponentials of the specific
linear combinations $H^i_{(1),n} + H^i_{(2),n}$ of the generators of
the \csa\ \gbo.}) characters $\chii_\Lambda(\tau)$ of \g\ into
(Virasoro specialized) characters $\chii_{\Lambda'}$ of \gP.
The modular transformation properties of the characters of \g\ and \gP\ imply
that the branching functions transform as
  \be  b_{\Lambda,\Lambda'}(-\Frac1\tau) = \sum_{M,M'}
  S_{\Lambda,M} S_{\Lambda',M'}^* \, b_{M,M'}(\tau) \, , \labl{bSb}
where the sum on $M$ and $M'$ is over all integrable highest weights of \g\
and \gP, \resp, at the relevant levels.

It is easy to check that for diagonal cosets the relation
  \be   (\lAb{\omD 0},\Hb)
  = (\lAbP{\omDP 0'}\eins,\HbP_\eins) + (\lAbP{\omDP 0'}\zwei,\HbP_\zwei)
  \labl{neurel}
holds. (For \gdc s one has to sum over the various ideals that make up \g\ and
\gp; then the analogous
relation takes the form $ \sumpe p (\lAb{\omD 0}\pp,\Hb_\pp)
  = \sumqe{p'} (\lAbP{\omDP 0'}\qq,\HbP_\qq)$.)
\futnote{A relation of this type can actually be derived in a much more general
setting for any automorphism that fulfills the \facprop.
The derivation, which is presented in appendix \ref{s.afac},
yields as a by-product the assertion that
the modular anomaly of branching functions is uniquely determined.}

The relation \erf{neurel} has two important consequences:
first, it leads to a conjugacy class selection rule.
This arises because $c_\lambda:=
(\lab\omdo,\lambdab)\mod\zet$ defines a conjugacy class of the \g-weight
$\lambda$; in \cft\ terms, this class is the monodromy charge \cite{scya}
of $\lambda$ \wrt the simple current, and the identity \erf{neurel} asserts
that
only those branching functions can be non-zero which correspond to
combinations of weights which have zero monodromy charge \wrt
all identification currents. Conversely, all selection
rules in a diagonal coset can be described
by identification currents which are all of the form $(J_\h,J_\h;J_\h)$, where
$J_\h$ can be any simple current of \h, because for diagonal cosets the
conjugacy class selection rules are precisely those which come from the
conjugacy class selection rules for the tensor products of \hb-\rep s.
 \futnot{ The fact that the form of (conjugacy class) selection rules does not
depend on the level is extremely restrictive and can be used for most
diagonal cosets to show that only selection rules of the form
$(J_\h,J_\h/J_\h)$
occur: in many cases, they are the only combinations of simple currents which
have integer spin for {\em all} levels.}
Therefore all selection rules can be implemented by strictly outer
automorphisms of \g\ and \gP\ that fulfill the \facprop. Also note that the
selection rule is an identity for conformal weights only modulo integers,
while the condition \erf{neurel} is valid exactly.

A second important consequence of the relation \erf{neurel} which follows
from the transformation  \erf{omvir} of the Virasoro algebra under strictly
outer automorphisms is that the coset Virasoro generators
$\Lc_m = L_m - \LP_m $ are invariant under $\omega$:
  \be  \bearll  \om(\Lc_m) \!\!&=\om(L_m)-\omP(\LP_m) \\[.3em]&
  = \llb L_m- \sumpe p (\lAb{\omD 0}\pp,\Hb_{\pp,m}) \lrb
  - \llb \LP_m - \sumqe{p'} (\lAbP{\omDP 0'}\qq,\HbP_{\qq,m}) \lrb
  = \Lc_m \, . \eear  \labl{omcvir}

\subsection{Identification of branching spaces and the coset chiral algebra}

Together with the $\omega$-twining property \erf{C'} of $\tauc$,
the invariance property \erf{omcvir} of the \cvira\ implies that
  \be  \tauc\,\Lc_n= \om(\Lc_n)\,\tauc= \Lc_n\,\tauc \,, \labl{tclc}
i.e.\ the map \tauc\ intertwines the action of the \cvira\
on the spaces \hll\ and \hlolo, i.e.\ we have $\tauc\!:\;\hll\to\hlolo$ with
  \be  [\tauc,\Lc_m]=0 \,   \labl{tclc2}
for all $m\iN\zet$.
(At this point it is worth while recalling our general convention that we
do not display the \rep s $R$ in which $\Lc_m$ acts explicitly; thus
equation \erf{tclc2} stands for the commutative diagram
  \be \begin{array}{rcl} \hll& \stackrel{\displaystyle\rll(\Lc_m)}
  {\mbox{------------------}\!\!\!\longrightarrow} & \hll\\[2 mm]
  {\tauc}\,\downarrow\ && \ \downarrow\,{\tauc}
  \\[2 mm] \hlolo \!\!\!  & \stackrel{\displaystyle\rlolo(\Lc_m)}
  {\mbox{------------------}\!\!\!\longrightarrow} & \!\hlolo \end{array}\ee
i.e.\ for\, $\tauc\circ\rll(\Lc_m)=\rlolo(\Lc_m) \circ \tauc$.)
This shows in particular that $\hll$ and $\hlolo$ carry isomorphic \rep s
of the coset Virasoro algebra. Thus
our result is a major step towards implementing field identification directly
on
the representation spaces
(up to now it was only known that the Virasoro specialized characters
of $\hll$ and $\hlolo$, i.e.\ the branching functions, are identical).

In order for the coset theory to be a fully consistent \cft, the intertwining
property of $\tauc$ must be valid not only for the \cvira, but for
all of the (maximally extended) chiral algebra \calwo\ of the coset theory. The
coset chiral algebra is commonly defined as the commutant of $\gP$ in
some suitable extension \ugtw\ of the universal enveloping algebra \ug\ of \g;
this extension must contain certain normal-ordered infinite sums of elements
of \ug, like those occurring in the Sugawara construction. Obviously, the
automorphism $\omega$ of \g, which extends to an automorphism of order $N$
of the enveloping algebra \ug, should also extend to an automorphism of
\ugtw\ of the same order. Such an extension of $\omega$
must, of course, respect the normal ordering prescription.

Let us denote by \calw\ the subspace of \ugtw\ which contains all elements that
commute with $\gP \subset \ugtw$. This subspace is in fact a subalgebra of
\ugtw\ which contains the coset Virasoro algebra. All elements of
$\calw$ have a well-defined action on branching spaces.
Moreover, the \facprop\ of $\omega$
implies that $\omega(\calw)\subseteq \calw$ for any automorphism $\omega$
that we need for the description of field identification; by an
analogous argument as in
the case of the endomorphism $\omP$ \erf1, $\omega$ therefore induces an
automorphism of order $N$ of \calw. Thus we can decompose \calw\ into
eigenspaces with respect to all automorphisms $\omega$ that fulfill the
\facprop.

The eigenspace $\calwoo$ of \calw\ that is left invariant under
all these automorphisms plays a particularly important role.
It is a subalgebra of \calw\ which according to \erf{omvir}
contains the coset Virasoro algebra. It contains all elements of $\calw$ that
are intertwined by $\tauc$. Hence, the implementation of field identification
by the maps $\tauc$ implies that $\calwoo$ rather than \calw\ is the
coset chiral algebra \calwo:
  \be \calwo  = \calwoo \, \, . \ee
Note that this imposes additional constraints on a
consistent coset chiral algebra, which have not been realized in the
literature so far.
In the next section we will see that precisely these constraints enforce
a consistent prescription for the resolution of field identification
fixed points.

\Sect{Fixed point characters and $S$-matrices}{fp}

We are now in a position to discuss the resolution of field identification
fixed points in generalized diagonal
coset \cfts. To this end we first recall that \Gid, the
{\em identification group}, is the abelian group generated by all
combinations of simple currents
of $\g$ and $\gP$ that describe selection rules of the theory. Recall that
for \gdc s all of them stem from strictly outer automorphisms that fulfill
the \facprop\ \erf1. A non-empty branching space is called a
{\em fixed point} if the relevant highest weights $\Lambda$ and $\Lambda'$
satisfy
  \be  \omt(\Lambda)=\Lambda\,,\qquad  \omtP(\LambdaP)=\LambdaP\,.  \labl{fix}
For diagonal cosets the fixed points are described by a combination of three
weights of \h\ that are fixed under a simple current automorphism $\omh$ of \h.

\subsection{Twining branching functions}

Recall that for any $\om\in\Gidp$ there is a linear map
  \be  \tauc:\quad \hll\to\hlolo \,; \labl{taucfix}
in case $\Lambda,\Lambda'$ corresponds to a fixed point of $\omega$, the map
$\tauc$ is even an endomorphism, and we can define the {\em\tbf} $\bllo$ as the
trace
  \be  \bllo(\tau) := \trhll\, \tauc \, \qc   \, . \labl{bllo}
In other words, the \tbf\ can be seen as the generating functional of
the trace of $\tauc$ on the subspaces having definite grade \wrt\ the
coset Virasoro algebra. They are thus generalized character-valued indices.
\futnot{here the qualification `generalized ' refers to the fact
that the map $\tauc$ does not necessarily have order two.}

These \tbf s turn out to be key ingredients in the description of fixed point
resolution.
In this section, we will therefore derive some of their properties using the
theory of \omchar s of \wzwts\ \cite{fusS3}. The {\em \ocha\/} $\chil$ of the
\ihwm\ \hl\ of \g\ \wrtt \soa\ $\om$ is defined as \cite{fusS3}
  \be  \chil(h):= \trhl \tauo\, \eE^{2\pi\ii R_\Lambda(h)}  \,.  \ee
In the present context we are mainly interested in the Virasoro specialized
(and anomaly-modified)
\ocha\ which for simplicity we denote by the same symbol, i.e.
  \be  \chil(\tau)= \trhl \tauo\, \eE^{2\pi\ii\tau (L_0-c/24))}  \,,
  \labl{chil}
and as usual we simply write $L_0$ in place of $R_\Lambda(L_0)$.
(In \cite{fusS3} only the case when \gb\ is simple has been described
explicitly; in the
general case, one has to take a suitable product of \omchar s of the various
summands of \g.) For notational convenience we define formally the \tbf\
$\bllo$ to be zero if $(\Lambda;\Lambda')$ is not a fixed point of $\omega$.

The main result of \cite{fusS3} was that the \omchar\ coincides with the
character of another \kma\ \gO, the so-called {\em\olie}. The \olie\
corresponding to \g\ and its diagram automorphism $\omega$ is obtained by
a simple prescription which corresponds to folding the \dyd\ of \g\ according
to the action of $\omD$.
Namely, the nodes of the \dyd\ of the \olie\ \gO\ correspond to the orbits
of $\omD$ on the \dyd\ of \g; the relative length of the corresponding simple
roots is given by the relative length of the orbits.
More precisely, the Cartan matrix $\AO$ of the \olie\ $\gO$ is given by the
formula
  \be \AOc ij := s_i \frac{N_i}N  \sum_{l=0}^{N-1} \Ac{\omdd li}j \,, \labl{AO}
where
  \be  s_i:=1- \sum_{l=1}^{N_i-1} \Ac{\omdd li}i  \,,  \labl{def.si}.
$N$ is the order of $\om$ and
$N_i$ the length of the $\omD$-orbit through $i$, and where
$[i],[j]$ take values in the set of $\omD$-orbits (or, alternatively, in a
label set which contains one chosen representative for each $\omD$-orbit).
All diagram automorphisms of \aff s satisfy $s_i\in\{1,2\}$ for all $i$,
except for the automorphism of order $N$ of $\g=A_{N-1}^{(1)}$
that rotates the \dyd, which has $s=0$ and leads to a trivial \tcha\
\cite{fusS3}.

Except for a special class of theories,
for which the \olie s are twisted \kma s,
these \olie s correspond precisely to the fixed point theories introduced
in \cite{scya5}.
It is a highly non-trivial property of the twining characters that, in case
$\omega$ is a diagram automorphism that corresponds to a simple current of \g,
the \omchar s form (up to a shift in the modular anomaly) a unitary \rep\ of
the twofold covering \slz\ of the modular group.

To be able to apply these results to the description of \tbf s, we will
show that $\bllo(\tau)$ arises in the decomposition of the \ocha s of \g\
\wrtt automorphism $\tauo$ into \ocha s of \gP\ \wrt $\tauoP$, i.e.\
they obey
  \be  \chil(\tau)= \sum_{\LambdaP} \bllo(\tau)\,\chilP(\tau) \,, \labl{cbc}
which also justifies the name `\tbf'.

To prove \erf{cbc}, we first note that (with the convention that the `trace'
is zero if $\tauo$ maps \hlPo\ to \hlPop\ with $\elp\ne\el$)
  \be  \chil(\tau)= \trhl \tauo\, q_{}^{L_0-C/24} = \sum_\el
  \trhlPo \tauo\, q_{}^{\LP_0-\CP/24} \qc \,,  \ee
because the \cvira\ commutes with $\LP_n$.
Owing to the fact that $\tauc$ and the coset Virasoro algebra commute
\erf{tclc2}, this can
be recast in the form
  \be  \chil(\tau)= \sum_\el \trhlP \tauoP\, q_{}^{\LP_0-\CP/24}\,
  \tauc\, \qc  \,,  \ee
or, rewriting the summation indices and using the fact that
non-vanishing contributions to the \rhs\ can only arise for $\taud\el=\el$, as
  \be  \bearll  \chil(\tau) \!\!&
  = \dsum_{M'} \dsum_{\scriptstyle\el\atop \scriptstyle\LambdaP(\el)=M'}
  \trhlP \tauoP\, q_{}^{\LP_0-\CP/24}\, \tauc\, \qc \\{}\\[-.5em]&
  = \dsum_{M'} \trhPM(\tauoP\, q_{}^{\LP_0-\CP/24}) \cdot
  \trhllM(\tauc\, \qc)  \,.  \eear\ee
By the defining relations of the \omchar s \erf{chil} and the \tbf s
\erf{bllo}, this is nothing but \erf{cbc}.

The formula \erf{cbc} means in particular that, analogously to the case of
ordinary branching functions, the \tbf s \bllo\ are computable from the
\tcha s $\chil$ and $\chilP$. Since the latter are equivalent to the
characters of the \olie s, the \tbf s can be computed as the ordinary
branching function
    \be  \bllo = \brev b^{}_{\brev\Lambda,\brev\Lambda'} \labl{brevb}
of the coset $\hO\oplus \hO / \hO$, where $\hO$ is the \olie\ of \h\
\wrt $\omh$.

Moreover, just as for the ordinary branching functions,
the \tbf s \bllo\ have nice modular transformation properties.
Given the modular transformation properties of the \omchar s of \g,
  \be  \chilastau = \sum_{\mu} \So\lambda\mu\, \chimtau \, ,  \ee
and of the \omchar s of \gP,
  \be  \chilastauP = \sum_{\muP} \SoP\lambdaP\muP\, \chimtauP  \,,  \ee
we find that
  \be  \bllostau = \sum_{(\mu;\muP)} \Sob\lambda\lambdaP\mu\muP\, \bmmotau
  \,. \labl{bsb}
Here we introduced
  \be  \Sob\lambda\lambdaP\mu\muP:= \So\lambda\mu\,(\SoP\lambdaP\muP)^*
  \,, \labl{Sob}
where $\So\lambda\mu$ \resp\ $\SoP\lambdaP\muP$ stands for the product of the
corresponding $S$-matrices of the \omchar s for all the ideals of \g\ \resp\
\gP. It would be rather difficult to compute the matrix elements
$\So\lambda\mu$ and $\SoP\lambdaP\muP$ from the definition \erf{chil} of
the \ocha s. However, using the results of \cite{fusS3} these matrices
can be easily computed as the $S$-matrices of the respective \olie s.
Also, the matrix
${\cal T}^{[\omega]}$ which describes the behaviour of the \tbf s under
$\tau\mapsto\tau+1$ is the diagonal unitary matrix obtained as the restriction
of the ordinary (untwined) $T$-matrix  $\CT$ to the fixed points.
This matrix is defined as the restriction to orbits of
non-vanishing branching functions of the product
$T^{\g}\otimes (T^{\g'})^*$ of $T$-matrices for the characters of $\g$ and
$\gP$.

We can summarize that the \tbf s are a set of functions which
can be described as the branching functions \erf{brevb} of the diagonal
coset of the \olie\ $\hO$ of \h. In particular they have nice modular
transformation properties; the corresponding $S$- and $T$-matrices are given
by the product of the $S$- and $T$-matrices of the \olie s.

\subsection{Identities for \tbf s and $S$-matrices}

Before describing the implementation of fixed point
resolution, we still need
two more relations. Namely, it can happen that a fixed point is left
fixed only by part of the identification group, while the rest of
the identification group induces field identifications. We would
like to identify the corresponding \tbf s and their \smat\ elements also
in this situation. In the present section we will derive relations for the
\tbf s (cf.\ \erf{sssb}) and their \smat\ (cf.\ \erf{sss}) which show
that this is indeed possible.

The first important observation we can make is that when performing
the field identification required for some simple current $J_1$, the
simple current of \g\ with \hw\ $J_2:= k_{(p)} \Lambda_{(\omD_2 0)}$
gives rise to a simple current of the \olie\ \gO. Namely,
simple currents correspond to so-called cominimal fundamental weights
$\Lambda_{(\omD_2 0)}$ of \g\
(except for the isolated simple current of $E_8$ at level two),
i.e.\ the associated Coxeter label has the value $a_{\omD_2 0} = 1$.
Because of the identity \cite{fusS3} $\aO_{[i]} = a_i/s_i$, it follows that
for the $\omD_1$-orbit $[\omD_2 0]$ we have
$\aO_{[\omD_2 0]}=1$ as well, which proves the observation. Note, however,
that the simple current $\JO_2$
of $\gO$ on which $J_2$ is projected can be trivial, i.e.\ the identity
primary field, as is definitely the case for the orbit of the current $J_1$
itself. Furthermore, the converse of the statement does not hold, i.e.\
it is not true that any simple current of $\gO$ arises in this way.
Namely, in case
there is an orbit with $s_i=2$, it can give rise to an additional simple
current of $\gO$ (as an example, take $\g=C^{(1)}_{2n+1}$ and its unique
non-trivial simple current, which leads to the \olie\ $\gO=
C^{(1)}_{n}$; the non-trivial simple current of \gO\ arises precisely from the
unique orbit of $\gO$ that has $s_i=2$).

Further, since the group of simple currents is abelian so that any two
diagram automorphisms commute (see \erf{ooo}), the action of $\JO_2$ on
the primary fields of $\gO$ reproduces the action of $J_2$ on $J_1$-orbits
of primary fields of $\g$. To describe coset theories, we have to combine
the corresponding statements for \g\ and \gP; in the case of fixed
points, the pair $\ilabel$
of weights of \g\ and \gP\ that characterizes a branching function
gets projected down to a pair $\ilabelO$ of weights of $\gO$ and $\gPO$.

To analyze the situation for the identification current $(J_\h,J_\h;J_\h)$,
we can assume that the \olie\ $\hO$ is again an untwisted affine \lie\ and that
there is again a coset \cft\ associated to the embedding \erf{coso}. In the
cases
(compare e.g.\ section \ref{s.bex}) for which the \olie\ is a twisted affine
algebra, the simple current has order 2 and the stabilizer is either trivial or
the whole identification group. It is therefore natural to look at the diagonal
embedding
  \be \hO \emb \hO \oplus \hO \, , \labl{coso}
where $\hO$ is the \olie\ of \h\ \wrt the simple current $J_\h$.\,
We claim that the fixed points of $\g/\gP$ which are allowed by the selection
rules are in one-to-one correspondence to the branching functions allowed
by the selection rules of the embedding \erf{coso}.

To compute the monodromy charge
  \be  \QO_2(\ilabelO) \equiv \hitO_\ilabelO - \hitO_{\JO_2\star\ilabelO}
  + \hitO_{\JO_2} \mod\zet  \ee
of the field labeled by $\ilabelO$ \wrt the simple current $\JO_2$, we note
that in the transition from $\g$ to $\gO$ the
conformal weights are shifted by a constant that does not depend on the
branching function \cite{fusS3}, so that we have
  \be \QO_2(\ilabelO) = h_\ilabel - h_{J_2\star\ilabel} + \hitO_{\JO_2}
  \mod\zet = \hitO_{\JO_2} \mod\zet \, . \labl{qo2}
Here in the last equality, we use that $J_2$ is an identification current,
which implies that $h_\ilabel-h_{J_2\star\ilabel}\iN\zet$. According to
the relation \erf{qo2},
the monodromy charge of all allowed fixed points is a constant that does not
depend on the specific fixed point.
Since the simple current $\JO_2$ describes a selection rule for the embedding
\erf{coso}, it has spin 0.\,\,%
\futnote{Note that in the case when $\hO$ is a twisted affine
\lie\ the identification current has order two so that fixed points
with different stabilizer do not exist.}
It follows that the right hand side of \erf{qo2} vanishes. Thus
we have derived that the monodromy charges of the projected
fixed points \wrtt projected simple currents are zero; in other words,
selection rules get projected down to selection rules.

Two immediate consequences of this statement are the following. First,
using \erf{sjs} we have
  \be  \CS^{[\omega_1]}_{\omzD\ilabel, \jlabel}
  = \CSO_{\JO_2\star \ilabelO, \jlabelO} = \CSO_{\ilabelO,\jlabelO}
  = \CS^{[\omega_1]}_{\ilabel, \jlabel} \, , \labl{sss}
i.e.\ $\CS^{[\omega_1]}_{\ilabel,\jlabel}$ does not depend on the choice of
representative of the $\omega_2$-orbit. Second, from the identity \erf{sss}
we conclude further that
  \be  \bearll  b^{[\omega_1]}_{\omzD\Lambda,\omzDP\lP} (-\Frac1\tau)
  \!\!&
  = \dsum_{\jlabel}\! \CS^{[\omega_1]}_{\omzD \ilabel, \jlabel}\,
  b^{[\omega_1]}_{\jlab}(\tau) \\{}\\[-.5em] &
  = \dsum_{\jlabel}\! \CS^{[\omega_1]}_{\ilabel, \jlabel}\,
  b^{[\omega_1]}_{\jlab} (\tau) = b^{[\omega_1]}_{\ilab} (-\Frac1\tau)
  \, . \eear \labl{sssb}
In other words, just as ordinary branching functions, the \tbf s
are identified as well.

\subsection{Resolution of fixed points}

As seen in section \ref{s.bs}, field identification requires that all
elements of the coset chiral algebra $\calwo$ have to commute with the map
$\tauo$ for all automorphisms
$\omega\in\Gid$. If a branching space \hll\ with branching function
\bll\ has a non-trivial stabilizer group
  \be  \Gilabel = \{\om\in\Gid \mid \omt\Lambda=\Lambda,\;\omtP\lP=\lP \} \,,
  \ee
then the corresponding
mappings $\tauo$ constitute endomorphisms of \hll. The fact that the action
of the coset algebra and the mappings $\tauo$ commute then implies
immediately that \hll\ carries a reducible \rep\ of the coset chiral
algebra $\calwo$, since already the eigenspaces \wrt $\tauo$ form modules
over $\calwo$. This way the implementation of field identification by the
maps $\tauo$ enforces fixed point resolution: one has to split the branching
spaces \hll\ into the eigenspaces \wrt the action of the stabilizer $\Gilabel$.

More precisely, let us consider any fixed branching space \hll, with
branching function \bll\
and stabilizer group $\Gilabel\subseteq\Gid$. Any character
  \be  \ch:\quad \Gilabel\to\complex  \labl{ch}
of the stabilizer group $\Gilabel$ gives rise to an eigenspace \hllch\ of \hll\
on which $\tauo$ has eigenvalue $\ch(\om)$, i.e.\ acts like $\ch(\om)$
times the identity map.
Just as the branching functions for the spaces \hll\ are denoted by
\bll, we denote the (Virasoro specialized) character of the eigenspace
\hllch\ by \bllch. By construction, we have
  \be  \bll(\tau) = \sum_{\ch\in\Gstabs} \bllch(\tau) \,, \labl{bb}
where $\Gilabels$ is the group of characters of $\Gilabel$. Since
the coefficients in the expansion of \bllch\ in powers of $q=\exp(2\pi\ii\tau)$
count the number of eigenstates at the relevant grade, they are
manifestly non-negative integers, and hence they are natural candidates
for the characters of the resolved fixed points.

Since \tauc\ acts on \hllch\ as the multiplication with $\ch(\om)$, it follows
that
  \be  \bllo = \sumGilabels \trhllch \qc \cdot \ch(\om)
  = \sumGilabels \ch(\om)\,\bllch \,. \labl{nset}
The orthogonality relation
  \be  \sumGilabel \ch_1^*(\om)\,\ch_2(\om)
  = |\Gilabel|\,\delta^{}_{\ch_1,\ch_2} \ee
of the characters of $\Gstab$ then implies that
  \be  \bllch = \Frac1{|\Gilabel|}\, \sumGilabel \!\!
  \ch^*(\om)\, \bllo \,. \labl{href}
Now recall that $\bllo$ is zero whenever $\omega\not\in \Gilabel$. It is
therefore consistent to extend $\ch\in\Gilabels$ to a function on the full
identification group $\Gid$
by setting $\ch(\omega) = 0$ for $\omega\not\in \Gilabel$.
(This extended function is {\em not} a character of $\Gid$.)
With this convention, which will simplify various formul\ae\ later on, we
can rewrite \erf{href} as
  \be  \bllch = \Frac1{|\Gilabel|}\, \sumGid \ch^*(\om)\, \bllo \,.
  \labl{sumchb}

At this point the following comments concerning field identification are
in order. As already seen in \erf{ooo}, diagram automorphisms
corresponding to simple currents
commute (the group of simple currents is abelian); this implies that the
maps $\tau_{\omega_1}$ and $\tau_{\omega_2}$ induced by any two
simple current automorphisms $\omega_1$ and $\omega_2$ commute as well.
Suppose now that
$\omega_2\not\in\Gilabel$ so that $\omega_2$ induces a field identification.
We then have
  \be  \Gilabel = \Gomilabelz\,, \ee
i.e.\ the stabilizers of all branching
functions that have to be identified are identical.
Moreover, $\tau_{\omega_2}$ respects the decomposition into
eigenspaces. Namely, if $v\iN\hllch$, i.e.\ $\tau_{\omega_1} v= \ch(\omega_1)
\, v$ for all $\omega_1\iN \Gilabel$, then also
  \be \tau_{\omega_1} (\tau_{\omega_2} v) = \tau_{\omega_2} \tau_{\omega_1} v
  = \ch(\omega_1)\,\tau_{\omega_2} v \, . \ee
This shows that $\tau_{\omega_2}$ implements in fact the field identification
between \hllch\ and ${\cal H}^{(\ch)}_{\oMT_2\Lambda,\omTP_2\Lambda'}$, and the
corresponding characters coincide. The characters of resolved fixed points
of the coset theory are therefore
  \be \chillch := \Frac1{|\GId|} \sum_{\omega\in\GId}\!
  b^{(\ch)}_{\omt\Lambda,\omtP\Lambda'}
  = \Frac1{|\GId|\cdot |\Gilabel| } \sum_{\ome,\omz\in\GId} \ch^*(\ome) \,
  b^{[\ome]}_{\omzD\Lambda,\omzDP\Lambda'} \, , \labl{crf}
or, inserting the identity \erf{sssb},
  \be \chillch = \Frac1{|\Gilabel|} \sum_{\om\in\GId} \ch^*(\om) \,
  b^{[\om]}_{\Lambda,\Lambda'} = \bllch \, . \labl{crF}
Thus the true characters of diagonal coset \cfts\ can be
expressed entirely in terms of branching functions and \tbf s.

\subsection{Computation of the \smat}

With the above results we are now also in a position to compute the \smat\
\SOb\ explicitly. To this end we denote by the pair $\ilabl$ of highest
weights in square brackets the orbit of $\ilabel$ \wrtt identification
group \Gid. Then we find
  \be \begin{array}{ll} \chillch(-\Frac1\tau) \!\!& =
  \Frac1{|\GId|\cdot |\Gilabel| }  \dsum_{\ome,\omz\in\GId} \ch^*(\ome)\,
  b^{[\ome]}_{\omzD\Lambda,\omzDP\Lambda'} (-\Frac1\tau) \\{}\\[-3mm]
  &=
  \Frac1{|\GId|\cdot |\Gilabel|\cdot |\Gjlabel|}
  \dsum_{\scriptstyle\ome,\omz,\omdr\atop\scriptstyle\in\GId}
  \dsum_{\jlabl} \ch^*(\ome)\, \CS^{[\ome]}_{(\omzD\Lambda;\omzDP\Lambda'),
  (\omdrD M;\omdrDP M')}\, b^{[\ome]}_{\omdrD M,\omdrDP M'}(\tau)\\{}\\[-3mm]
  &=
  \Frac1{|\GId|\cdot |\Gilabel|\cdot |\Gjlabel|}
  \dsum_{\scriptstyle\ome,\omz,\omdr\atop\scriptstyle\in\GId}
  \dsum_{\jlabl} \ch^*(\ome)\, \CS^{[\ome]}_{(\omzD\Lambda;\omzDP\Lambda'),
  (\omdrD M;\omdrDP M')}
  \\&\hsp{19}\times\! \dsum_{\cht\in\Gjlabels}\!\!\! \cht(\ome)\,
  b^{(\cht)}_{\omdrD M,\omdrDP M'}(\tau) \\{}\\[-3mm]
  &=
  \Frac1{|\Gilabel|\cdot |\Gjlabel| } \dsum_{\jlabl}
  \dsum_{\cht\in\Gjlabels} \dsum_{\ome,\omdr\in\GId}\!
  \ch^*(\ome)\, \CS^{[\ome]}_{\ilabel,\jlabel}
  \cht(\ome)\, b^{(\cht)}_{\omdrD M,\omdrDP M'}(\tau) \\{}\\[-3mm]
  &=
  \Frac{|\GId|}{|\Gilabel|\cdot |\Gjlabel|} \dsum_{\jlabl}
  \dsum_{\cht\in\Gjlabels} \!\llb \sum_{\ome\in\GId}\!\!
  \ch^*(\ome)\, \CS^{[\ome]}_{\ilabel,\jlabel}\, \cht(\ome) \lrb \,
  \chimmcht(\tau)
  \,. \eear \ee
{}From this formula we read off that the actual \smat\ for a coset \cft\ with
fixed points is
  \be  \Scc\Lambda\LambdaP M{M'} :=
  \frac{|\GId|}{|\Gilabel|\cdot |\Gjlabel| } \sum_{\om\in\GId}
  \ch^*(\om)\, \CS^{[\om]}_{\ilabl,\jlabl} \, \cht(\om) \, . \labl{Scc}
Note that here $\ch\iN\Gilabels$ and $\cht\iN\Gjlabels$, so that
according to our conventions, the summation is effectively only over the
subgroup
$\Gilabel\cap\Gjlabel$ of $\Gid$.

In appendix \ref{s.scheck} we will check that the matrix \erf{Scc}
is symmetric and unitary and
that its square gives a permutation of order two on the primary fields of the
coset \cft. Moreover, as already mentioned we obtain a diagonal unitary
matrix $\CT$ by using the conformal weights of the branching functions.
Thus the different fields
into which a fixed point is resolved have the same $T$-matrix elements
(although
due to the character modifications
their conformal weights might differ by integers):
  \be \CT_{(\ilabl,\ch),(\jlabl,\cht)} = \delta_{\ilabl,\jlabl}\,
  \delta_{\ch,\cht}\, T_{\ilabel} \, . \labl{tdef}
In the appendix we will also check $\CS$ and $\CT$ generate a unitary \rep\
of \slz. Furthermore, in all cases we have checked
explicitly, the matrix $\CS$ gives rise via the Verlinde formula
to non-negative integral fusion coefficients. \futnot{We should prove this}

\bigskip
As an illustration, consider the case where the identification group \Gid\ is
isomorphic to the cyclic group $\zet_N$ with $N$ prime. Then the identification
group $\Gid$ has $N$ characters $\ch=\ch_k$, $k=\otoNm$, acting on $n\iN\zet_N$
as
  \be  \ch_k(n) = \zeta^{kn}  \,,  \ee
where
  \be  \zeta:=\exp\LLb\frac{2\pi\ii}N\LRb  \labl{zeta}
is a primitive $N$th root of unity.
Then for any automorphism $\om\ne\id$ in \Gid\ the matrix $\SNb$ is the same
for all $n=\onetoNm$,
  \be  \Snb\lambda\lambdaP\mu\muP =: \Sf\lambda\lambdaP\mu\muP  \labl{Sf}
for $n\ne0$, while of course
  \be  \Sno\lambda\lambdaP\mu\muP = \Soo\lambda\lambdaP\mu\muP
  \equiv \Sm\lambda\mu\,(\SP\lambdaP\muP)^* \labl{Soo}
is the \smat\ describing the modular behavior of the ordinary branching
functions.

If either $\ilabel$ or $\jlabel$ is not a fixed point, the sum in \erf{Scc}
has only one term, corresponding to the identity element of $\Gid$, and the
\smat\ is a multiple of the ordinary \smat\ element for the corresponding
branching functions. More interesting is the case when both branching
functions are fixed points; then we have
  \be  \bearll  \Skl\lambda\lambdaP\mu\muP \!\!& =
  \Frac1N \dsum_{n=0}^{N-1} \ch_k^*(n)\, \Snb\lambda\lambdaP\mu\muP\,
  \ch^{}_l(n) = \Frac1N \dsum_{n=0}^{N-1} \zeta^{n(l-k)}\,
  \Snb\lambda\lambdaP\mu\muP\, \\{}\\[-.5em]&
  = \Frac1N \Soo\lambda\lambdaP\mu\muP + \Frac1N \Sf\lambda\lambdaP\mu\muP
  \cdot \dsum_{n=1}^{N-1} \zeta^{n(l-k)} \\{}\\[-.5em]&
  = \left\{ \bearll
    \Frac1N\, \Soo\lambda\lambdaP\mu\muP +
(1-\Frac1N)\,\Sf\lambda\lambdaP\mu\muP
    & {\rm for}\ l=k \,, \\[.8em]
    \Frac1N\, \Soo\lambda\lambdaP\mu\muP -\Frac1N\,\Sf\lambda\lambdaP\mu\muP
    & {\rm for}\ l\ne k \,. \eear \right.
  \eear \ee
This is precisely the expression for the \smat\ that has been conjectured
in \cite{scya5}.

\Sect{The diagonal $B_n$ embeddings}{bex}

To illustrate our results presented in the previous sections we study now
the diagonal coset theories based on
  \be  \g= (B_{n+1}^{(1)})_1^{} \oplus (B_{n+1}^{(1)})_1^{} \,, \qquad
  \gP=(B_{n+1}^{(1)})_2^{} \labl{bbb}
in some detail.
In these cosets, we have the identification current $(J,J;J)$, where $J$
denotes
the unique non-trivial simple current of $B_n^{(1)}$.

The coset theories \erf{bbb} have Virasoro central charge $\cc=1$,
so that we can determine their characters also via a different route, namely
by constructing them out of a free boson. The $c=1$ \cfts\
have been classified, and looking at the spectrum, it turns out that the
series $SO(N)_1 \oplus SO(N)_1 / SO(N)_2$ can be identified with
$\Zbf_2$ orbifolds of the $U(1)$ theory with $4N$ primary fields.
The comparison of the known character formul\ae\ for the latter theories
to the character modifications implied by our prescription provides us
with a rather non-trivial consistency check.
The coset description of these theories has fixed point problems
only for odd $N$, and in \erf{bbb} we have set $N=2n+3$.

An analysis of these cosets has already been presented in \cite{scya5}.
The main results of that work can be summarized as follows. The set of
conformal weights of the branching functions was shown to be equal to the
set of conformal weights of the non-unitary minimal Virasoro models
with $(p,q)=(2,N)$. Hence, according to the
conjecture formulated in \cite{scya5}, the characters of those models are
natural candidates for the character modifications. This conjecture
was verified explicitly by computing the relevant orbifold
characters, and showing that the difference between certain pairs
of them is indeed equal to the minimal model characters. \\[.2em]
Our new results improve the description of these coset theories
in two important ways: \vspace{-1mm}
\begin{quote}
1.\, We can {\it derive} the character modifications, rather than
conjecturing them.\\[.4em]
2.\, We can extend the analysis in principle to the entire series\\[.4em]
\mbox{}~~~ $B(n+1,k,l):=(B_{n+1}^{(1)})_k^{}\oplus (B_{n+1}^{(1)})_l^{}
/(B_{n+1}^{(1)})_{k+l}^{}$, \,for arbitrary $k$ and $l$.
\end{quote}
The reason why the procedure of \cite{scya5} was limited to $k=l=1$ is that
except for $k=l=1$
the fixed point conformal weights could not be identified with the spectrum
of any known \cft. Now according to the results of \cite{fusS3} the orbit
Lie algebra of $(B_{n+1}^{(1)})_k^{}$ is given by the twisted affine \lie\
$(\tilde B_{n}^{(2)})_k^{}$.
The results of the present paper then imply that the character modifications
for $B(n+1,k,l)$ must be equal to branching functions of
 \futnote{Although the notation used here is the same as the one we employ
for coset \cfts, we are not proposing here a definition of coset theories
of twisted affine algebras, but only of branching functions for the
embedding of the algebras.}
$(\tilde B_{n}^{(2)})_k^{} \oplus(\tilde B_{n}^{(2)})_l^{}
/ (\tilde B_{n}^{(2)})_{k+l}^{} =: \tilde B(n,k,l)$.
While this is true for arbitrary $k$ and $l$, it is only for $k=l=1$ that
we have an explicit check.

The coset theories $B(n+1,k,l)$ have an identification current $(J,J ; J)$
which satisfies the \facprop. For $k=l=1$  the fixed points are
the combinations  $(s,s ; r)$, where $s$ and $r\in\{1,2,\ldots,n+1\}$
denote the primary fields that carry the fundamental
spinor representation of $B_{n+1}$ and the
antisymmetric tensor representations, respectively.
As was shown in \cite{scya5}, after fixed point resolution the combination
$(s,s;r)$ is resolved into two primary fields of the orbifold conformal field
theory. Each of these two representations corresponds in its turn
to two representations of the $U(1)$ theory. This can be described
diagrammatically as follows:
  \be  (s,s;r)\stackrel{{\rm FP}}{\mbox{---}\!\!\!\to}
  \left\{ \begin{array}{ll} |q| = N-2r & \stackrel{{\rm
D}}{\mbox{---}\!\!\!\to}
  \left\{ \begin{array}{ll} q=N-2r\,, \\ q=-N+2r\,, \end{array} \right.
  \\{}\\[-.3em]
  |q| = N+2r & \stackrel{{\rm D}}{\mbox{---}\!\!\!\to}
  \left\{ \begin{array}{ll} q=N+2r\,, \\ q=-N-2r\,.
  \end{array} \right. \end{array} \right. \labl{Split}
Here the first step represents fixed point resolution of the coset theory,
whereas
the second step represents the extension of the orbifold chiral algebra
by a spin-1 simple current, which becomes the $U(1)$-current of
the theory on the circle.
The primary fields of the $U(1)$ theory are uniquely identified by
their $U(1)$-charge $q$. The orbifold procedure identifies opposite charges
and projects out the $U(1)$ current, but in the untwisted sector
the absolute value of the charge can still be used to label the fields.
Since opposite charges are identified in orbifolding, it follows that in the
inverse process of extending the orbifold chiral algebra the corresponding
fields are fixed points of the spin-1 simple current.

It is extremely important to be aware of the difference between the
two steps in the diagram \erf{Split}. The first step represents fixed
point resolution in a coset theory, and hence involves character
modifications, while the second represents fixed point resolution
in a `D-type' modular invariant, which does not involve any
character modification. Hence we have
  \be  b^{}_{s,s;r}= \chii^{\rm orb.}_{N-2r} + \chii^{\rm orb.}_{N+2r} \ee
and
  \be  \chii^{\rm orb.}_{N-2r} = \chii^{\rm circ.}_{N-2r}
  = \chii^{\rm  circ.}_{-N+2r} \,, \qquad
  \chii^{\rm orb.}_{N+2r} = \chii^{\rm circ.}_{N+2r}
  = \chii^{\rm circ.}_{-N-2r}\,,  \ee
where $\chii^{\rm orb.}_q$ and $\chii^{\rm circ.}_q$ are the Virasoro
specialized characters
of the orbifold and of the $U(1)$ theory, respectively; the latter are just
ratios of $\theta$-functions for the circle theory and the $\eta$-function.
The character modification can thus be written as
  \be  \elta^{}_{s,s;r}= \chii^{\rm circ.}_{N-2r}- \chii^{\rm circ.}_{N+2r}
  \,, \ee
where the overall sign was chosen in such a way that all $\elta$'s are
positive. We have now expressed the character modifications in terms
of known functions, which are ratios of $\theta$- and $\eta$-functions.
It was shown in \cite{scya5} that $\elta_{s,s;r}= \chii_{r,1}^{2,N}$,
where $\chii^{p,q}_{r,s}$ denotes a character of a Virasoro minimal model.

Now we turn to the description of these character modifications in terms
of branching functions of twisted Kac-Moody algebras. A fixed point
of the simple current of $B_{n+1}^{(1)}$ has Dynkin labels
$(a_0,a_1,a_2,\ldots,a_{n+1})$ with $a_0=a_1$,
where the first label refers to the extended simple
root. The twining character of the simple current automorphism was shown
in \cite{fusS3} to be equal to the $\tilde B_n^{(2)}$-character for
the highest weight with Dynkin labels
$(a_1,a_2,\ldots,a_{n+1})$, where the first label corresponds to the longest
simple root and the last one to the shortest. The unique fixed point $s$
of $B_{n+1}^{(1)}$ at level 1 has highest weight $(0,0,0,\ldots,0,1)$; it
is thus related to the $\tilde B_n^{(2)}$-representation with Dynkin
labels $(0,\ldots,0,1)$, which we will denote by $\sO$. At level 2 of
$B_{n+1}^{(1)}$ there are $n+1$ fixed points, which we denote by
$r\in\{1,2,\ldots,n+1\}$, with highest weight $(1,1,0,\ldots,0)$ for $r=1$,
$(0,0,0,\ldots,0,1,0,\dots,0)$ (with $a_{r+1}=1$) for $1\leq r\leq n$,
and $(0,0,\ldots,0,2)$ for $r=n+1$. They
are mapped respectively to the $\tilde B_n^{(2)}$-representations with
highest weights $(1,0,\ldots,0)$, $(0,\ldots,0,1,0,\ldots,0)$ and
$(0,\ldots,0,2)$, to which we will refer by $\rO$. The identity we wish
to check is then
  \be  \Llb  \chii_\sO^{\tilde B_{n,1}^{(2)}} \Lrb ^2_{}
  = \sum_{r=1}^{n+1} \elta^{}_{s,s;r}\,
  \chii_\rO^{\tilde B_{n,2}^{(2)}}\,.   \labl{postulate}

The characters for $B_{m}^{(1)}$ and $\tilde B_m^{(2)}$ can be written down
explicitly by choosing an orthonormal basis $e_1, \ldots,e_m$ in
the weight space of the horizontal $B_m$ Lie algebra. In this basis the
roots of $B_m$ are $\pm e_i \pm e_j$ with $ 1 \leq i < j \leq m$ and
$\pm e_i$ with $1 \leq i \leq m$.
The simple roots of the horizontal algebra are $e_1-e_2,\; e_2-e_3,\; \ldots,\;
e_{m-1}-e_m$ and $e_m$, and the extended simple root is $-e_1-e_2$ for
$B_m^{(1)}$ and $-2e_1$ for $\tilde B_m^{(2)}$.
The Weyl\hy Kac character formula yields in this basis
  \be  \chii_{\vec w,k} = q^{\delta(\vec w,k)}\, {{\calS}(\vec w,k) \over
  {\calS}(0,0)}\,, \ee
 \futnot{instead of $\calS$, this was $\cal S$; but we use the latter notation
already for the coset \smat}
where $\delta$ is the `modular anomaly' (which equals $h-c/24$ in a
conformal field theory), and
  \be  {\calS}(\vec w,k):=\sum_{\pi \in S_m} \epsilon({\pi}) \!\!
  \sum_{\scriptstyle n_i\in\zet\atop\scriptstyle\sum n_i = 0 \mod 2 }\prod_i
  \sin\lLb[w_i+\rho_i + n_i (k+\gv) ]\theta_{\pi(i)} \lRb\,
  q^{\onehalf(k+g_{}^\Vee)\vec n^2 + (\vec w + \vec \rho)\cdot \vec n}
  \labl{bmk}
for $B_{m,k}$. Here $\vec w$ is a highest weight of the
Lie algebra, $w_i$ are its components in the orthonormal basis, $k$ is
the level, $\vec\rho$ the Weyl vector (whose components in the
horizontal basis read $\vec\rho=({2n-1\over 2},
{2n-3\over2},\ldots,{1\over2})$) and $\gv$ the dual Coxeter number,
$\gv=2m-1$.
The variables $\theta_i$ correspond to
the generators of the Cartan subalgebra of $B_m^{(1)}$. As we are
only interested in Virasoro characters here,
we will eventually set them to zero, but this can only be done after
canceling common factors between the denominator and the numerator
of the character formula. The branching function is in fact independent
of the variables   $\theta_i$.

For the twisted affine \lie\
$(\tilde B_m^{(2)})_k$ almost the same formula as \erf{bmk} holds, namely
  \be  \tilde{\calS}(\vec w,k):=\sum_{\pi\in S_m}
\epsilon({\pi})\sum_{n_i\in\zet}
  \prod_i \sin\lLb[w_i+\rho_i + n_i (k+\tildegv)]\theta_\pi(i) \lRb\,
  q^{\onehalf(k+\tildegv)\vec n^2 + (\vec w + \vec \rho)\cdot \vec n}\,, \labl"
the only modifications being that now $\tildegv=2m+1$ and that the second sum
is over {\it all} integer-valued vectors $\vec n$. The vector $\vec w$ is
obtained in this case by writing the $\tilde B_m^{(2)}$ Dynkin labels
as $(b_0,b_1\ldots,b_m)$ and regarding $b_1,\ldots,b_m$ as
$B_m$ Dynkin labels. (To apply the formula \erf", the resulting highest
weight of $B_m$ must then be converted to the orthogonal basis.)

The modular anomaly for $B_m^{(1)}$ is
  \be  \delta_{B}(\vec w,k) = {\half(\Lambda+2\rho, \Lambda)_B^{} \over k+2m-1}
  -{1\over 24 }\,{m(2m+1)k\over k+2m-1} \,. \ee
For $\tilde B_m^{(2)}$ we find \futnot{The last term disagrees with Kac,
but cancel for the coset theory under consideration. \\
This footnote regarding a comparison with Kac should be reconsidered}
  \be  \delta_{\tilde B}(\vec w,k) = {\half(\Lambda + 2\rho, \Lambda)_B^{}
  \over k+2m+1}
  - {1\over 24}\, { m(2m-1) k\over k+2m+1} -{km\over 12} \,. \ee
Using these results one may check that for any fixed point
the branching functions of $B(n+1,k,l)$ have the same modular anomaly
as the corresponding ones of $\tilde B(n,k,l)$. This implies that for any
fixed point the branching function and the character modifications have
the same modular anomaly, so that all characters receive corrections to
their leading term. It also implies that we may drop the modular anomaly terms
when checking \erf{postulate}, since they cancel.

Although we have no general proof of the identity \erf{postulate}, the
character formul\ae\
are explicit enough to allow partial verification. In particular we have
checked
that for $n=1$ and $n=2$ the first 40 terms in the expansion in $q$ agree,
and we have verified the identity for all $n$ in leading and
next-to-leading order in $q$.
The leading order check is straightforward. Note that the $\tilde B_n^{(2)}$
character formula has as its leading term the dimension of the
representation of the horizontal $B_n$ algebra (as it does for the
untwisted algebra). Since on the left-hand side of \erf{postulate} the ground
state
is a spinor and on the right-hand side they are antisymmetric tensors
of rank $r-1$, the leading order of \erf{postulate} amounts to the condition
  \be  \left[ 2^{n}\right]^2 = \sum_{r=0}^n {2n+1 \choose r} \,, \ee
which is indeed satisfied.

\sect{Summary and outlook}

In this article, we have solved the problem of resolving field identification
fixed points of diagonal coset \cfts, and of fixed points in the even larger
class of \gdc s. To this end we have implemented both field identification
and fixed point resolution directly on the branching spaces. Technically,
this was achieved by the use of automorphisms of the \dyd\ which fulfill the
\facprop\ \erf1. They give rise to
$\omega$-twining maps $\tauc$ on the branching spaces.

One important insight that we gain from this construction is that the
coset chiral algebra $\calwo$ is {\em not} the commutant of the subalgebra
$\gP$ in (some closure of) the universal enveloping algebra of \ug, but
rather a subalgebra of it, namely the subalgebra that is fixed by all
simple current automorphisms of the \dyd\ that fulfill the \facprop.

We have seen that precisely this property enforces fixed point resolution:
on the fixed points, i.e.\ those branching spaces on which the identification
group acts non-freely, the action of the chiral algebra cannot change
the eigenvalue of eigenvectors \wrtt action of the stabilizer. Therefore we
must split these branching spaces
into the various eigenspaces \wrt that action.
The recently developed theory of \omchar s and \olie s shows that this
splitting is governed by a natural mathematical structure
which allows for explicit calculations. In particular, we have derived
expressions for both the characters and the \smat\ of the coset theory in terms
of the corresponding data of the \olie s and of group characters of the
identification group. Finally, we have made this resolution procedure
explicit for a specific example of cosets describing the $\zet_2$
orbifolds of a compactified free boson.

Our results can be extended in several directions. First of all,
one would also like to describe more general cosets than the generalized
diagonal ones. However, we do not know any other examples of coset \cfts\
where the \facprop\ holds. This implies that in the general case, one
cannot restrict oneself to strictly outer automorphisms $\omega, \omega'$ on
\g\ and \gP. Rather one has to allow for more general outer automorphisms
of \g\ and \gP, which are the product of a strictly outer automorphism
and some compensating inner automorphism corresponding to an element
of the Weyl group. A candidate for such automorphisms is the
spectral flow described in \cite{levw}.
However, we expect that most features we
encountered in the case of \gdc s will also be present in the general
case of arbitrary embeddings $\gP\emb\g$; in particular, we conjecture
that the formula \erf{Scc} for the \smat\ of the coset \cft\ will still be
valid.

Taking into account inner automorphisms will definitely be necessary for
the description of those identification currents which act non-trivially only
in the subalgebra $\gP$; these `$(1;J)$-currents' are expected to present
particular problems. In this context, it is interesting to remark that
the so-called maverick cosets \cite{dujo}, i.e.\ those cosets which
possess further selection rules in addition to the conjugacy class
selection rules, always have such identification
currents, and that these currents have fixed points. There are, however,
examples of coset \cfts\ which also fulfill these criteria, but which are
nevertheless not maverick.

As a side remark, we mention that
in addition to the maverick cosets found in \cite{dujo2}, there is at
least one other maverick coset, given by $\g=E_8$ and $\gP=A_1\oplus E_7$, with
all algebras at level 2.
 \futnote{Thus the A-D-E classification of maverick cosets that was conjectured
in \cite{dujo2} is not complete.}
 (The analogous embedding at level 1
is a conformal embedding, just as for the other maverick cosets). It has as
an identification
current $(1;J,J)$, where $J$ stands for the unique non-trivial simple current
of $A_1$ respectively $E_7$; this identification current has fixed points.
The branching rules for this coset (see \cite[\S4.3, example (j)]{kawa})
show that there is an additional non-vanishing branching function,
corresponding
to $(\Lambda_{(7)}; \Lambda_{(0)} + \Lambda_{(1)}, \Lambda_{(7)})$, that has
conformal weight zero and therefore should be identified with the vacuum of the
(putative) coset theory. However, the weights involved are not on
any simple current orbit of the vacuum.

Finally, we want to point out that there are still more open conceptual
problems in the description of the coset construction, which are
largely independent of the problem of field identification and fixed point
resolution. First of all, one should show that there is a suitable extension
$\ugtw$ of the universal enveloping algebra $\ug$ such that the candidate
coset chiral algebra \calwo\ fulfills all requirements for a chiral algebra
of a \cft.
Moreover, in order to prove that \calwo\ is indeed the -- maximally extended --
chiral algebra of the coset theory, one must, in the absence of
field identification fixed points, also show that the branching
spaces are {\em irreducible} modules of \calwo, and that each isomorphism
class of irreducible modules appears precisely once. (Note that for
$\cc\geq1$ the branching
spaces are highly reducible as modules over the \cvira.) In the presence of
field identification fixed points, one has to prove the same statements for
the eigenspaces of the maps $\tauo$.

\newpage
\ASect{The \facprop\ and the uniqueness of modular anomalies}{afac}

In this appendix we drop the assumption that the embedding $\gP\emb\g$
describes a \gdc. Whenever the embedding $\gP\emb\g$ preserves the triangular
decomposition, the restriction $\omP$ of a strictly outer automorphism of \g\
which satisfies the \facprop\
is again a strictly outer automorphism of \gP. Namely, first, as a
restriction of a homomorphism, $\omP$ is a homomorphism of \gP. Second, from
$\om^N=\id$ it follows that also $\omP^N=\id$; this implies that $\omP$ is
surjective and injective, and hence an automorphism of \gP; moreover, the order
\NP\ of $\omP$ divides the order $N$ of $\om$.
Since both the embedding $\gP\hookrightarrow\g$ and the automorphism
$\omega$ \erf{omz} respect the triangular decomposition of \g, we have
$\omP(\gzP)=\om(\gzP)\subseteq \om(\gz)\subseteq\gz$ \fortria, while
$\omP(\gzP)
\subseteq\gP$ by the \facprop\ of $\omP$. Together it follows
that $\omP(\gzP)\subseteq\gz\cap\gP=\gzP$, and hence $\omP$ is a strictly
outer automorphism of \gP.

It can also be shown that any strictly outer automorphism
satisfying the \facprop\ leads to a selection rule of the form \erf{neurel}
and leaves the generators of the coset Virasoro algebra invariant.
The proof of this assertion has some interesting by-products;
it proceeds as follows.
{}From the definition \erf{39} of the branching function it is apparent that
(up to an overall prefactor $q^{h_\Lambda-h_{\Lambda'} - \ccs/24}$, where
$h$ denotes the conformal weight) the coefficient $d_n$ of the power $q^n$ of
the branching function counts how many \hwv s of \gP\ with highest weight
$\Lambda'$ occur at grade $n$ in the \g-module with highest weight $\Lambda$.
Now the \hwv s of \gP\ in \hl\ with highest weight $\Lambda'$
are in one-to-one correspondence to \hwv s of \gP\ in \hlo\ with highest weight
$\omT(\Lambda)$ at the same grade of the module.
This implies that the associated branching functions are related by
  \be  b_{\omt\Lambda,\omtP\Lambda'}(\tau)
  = \eE^{2\pi\ii d_{\Lambda,\Lambda'} \tau} \bll(\tau) \, , \labl{ess}
where
  $    d_{\Lambda,\Lambda'} := h_{\oMT \Lambda}-h_{\Lambda}-h_{\oMT' \Lambda'}
  + h_{\Lambda'} $.
In particular, $b_{\oMT\Lambda,\oMT'\Lambda'}$ vanishes if and only if
$b_{\Lambda,\Lambda'}$ vanishes. Also, from \erf{bSb} we learn that
  \be b_{\oMT\Lambda,\oMT'\Lambda'}(-\Frac1\tau) = \sum_{M,M'}
  S_{\oMT\Lambda,M} S_{\oMT'\Lambda',M'}^* \, b_{M,M'}(\tau) \, . \ee
On the other hand, using \erf{ess}, we see that
  \be b_{\oMT\Lambda,\oMT'\Lambda'}(-\Frac1\tau) =
  \eE^{-2\pi\ii d_{\Lambda,\Lambda'} /\tau}  \sum_{M,M'}
  S_{\Lambda,M} S_{\Lambda',M'}^* \, b_{M,M'}(\tau) \, . \ee
Let us assume that $\bll$ does not vanish; then by comparing the two
results we obtain the identity
  \be \eE^{-2\pi\ii d_{\Lambda,\Lambda'} /\tau} =
  \frac{ \sum_{M,M'} S_{\oMT\Lambda,M} S_{\oMT'\Lambda',M'}^* b_{M,M'}(\tau)}
  {\sum_{M,M'} S_{\Lambda,M} S_{\Lambda',M'}^* b_{M,M'}(\tau) }\labl{contr}
between two functions on the complex upper half plane $H_+$
(${\rm Im}\,\tau >0$).

Since both $\g$ and $\gP$ have only a finite number of primary fields, we
can define $M$ as the smallest common denominator of the conformal weights
of all branching functions and rewrite \erf{contr} in terms of the variable
$x\! := \eE^{2\pi\ii\tau/M}$.
Branching functions are holomorphic functions on $H_+$
\cite[\S 1.6]{kawa}, and hence the right hand side of \erf{contr} converges to
a meromorphic function on $H_+$. Manifestly, we can also interpret it as a
meromorphic function in $x$ for $|x| <1$. After multiplying both numerator and
denominator of the right hand side with an appropriate power of $x$ to get rid
of negative powers of $x$ which stem from terms of the form $q^{-c/24}$
in the characters, we can assume that both contain only positive powers of $x$.

The left hand side of \erf{contr} is a priori defined not on $x$-space, but
rather
on an infinite covering of it. If both sides of \erf{contr} are identical, then
also this side has to give rise to a function in $x$ and we do not have to
worry about branch cuts any more. Since the left hand side of \erf{contr} tends
to $1$ for ${\rm Im}\, \tau\to \infty$, $x=0$ has to be a regular point for
both
sides, and hence also the derivatives of the left side \wrt $x$ have to exist
at $x=0$. However, the first derivative of the left side
diverges for $x\to1$, unless $d_{\Lambda,\Lambda'}$ vanishes.
\futnot{We have:
\def\dd{\tilde d}
$\tau = \frac1{2\pi\ii}M \ln x\,$, hence the lhs is
$\eE^{-2\pi\ii d \frac1\tau} = \eE^{4\pi^2 d/M \frac1{\ln x}}
= \eE^{\dd /\ln x}\,$. The first derivative of this function is
$- \frac{\dd \eE^{\dd /\ln x}}{(\ln x)^2 x} = -
\dd \eE^{\dd/\ln x} \frac1{(\ln x)^2 x} \,$.
The first two factors tend to $\dd$ for $x\to 0$, while the numerator tends to
$0$, (`The logarithm tends to $-\infty$ slower than any power'.) Hence the
derivative does not a exist: this is a contradiction. The only way to
avoid the contradiction is by $\dd=d=0$.}

Thus we have $d_{\Lambda,\Lambda'}=0$. {}From \erf{ess} we then learn that
  \be  b_{\oMT\Lambda,\oMT'\Lambda'} \equiv  b_{\Lambda,\Lambda'} \labl{blb}
for all pairs $(\Lambda, \Lambda')$ of integrable \hw s. In other words,
the eigenvalue equation $\Lc_0 v_\lambda = h v_\lambda$, where $v_\lambda$ is
an element of any weight space in any \irmod, implies that
$\Lc_0 \tauo v_\lambda = h \tauo v_\lambda$ as well.

The argument we just presented also gives some further important insight.
Suppose one is given a set of functions of the modular parameter $\tau$ which
transform under the transformation $\tau\mapsto-1/\tau$ to linear combinations
of these same functions, like e.g.\ the characters of some rational \cft\ or,
as
in the case of our interest, the branching functions. Our argument then
shows that, whenever
two such functions differ only by a factor which is some power of $q$,
they must in fact be identical. In the case of a rational \cft, this implies
that if the modules of two primary fields have the same number of states
at any degree, then the two primary fields must have the same
conformal weight.

With the help of the relation \erf{blb} the promised results are
recovered as follows. Taking into account the transformation \erf{omvir}
of the Virasoro algebra under strictly outer automorphisms
and its analogue for \gP, we can make the most general ansatz
  \be \omega^{-1}(\Lc_0) = \Lc_0 - (v,H_0) + \sumpe p \xi_p K_{(p)} \,  \ee
for $\omega^{-1}(\Lc_0)$, where $K_{(p)}$ are the central elements for
the various ideals of \g\ and the sum is over these ideals.
Using the $\omega$-twining property \erf{C'} of $\tauo$, we then see that
for any vector $v_\lambda$ with weight $\lambda$ in any \hwm\ of \g\ we have
  \be  \begin{array}{ll} h\,\tauo v_\lambda \!\!&
  = \Lc_0 \tauo v_\lambda = \tauo\omega^{-1}(\Lc_0) v_\lambda
  = \tauo(\Lc_0 - (v,H_0) + \sum_p \xi_p K_{(p)})\, v_\lambda \\[2.1mm]
  &= (h-(v,\lambda) + \sum_p \xi_p k_{(p)})\, \tauo v_\lambda \,. \eear\ee
Since this equation has to be valid at any combinations of levels $k_{(p)}$
and for any weight $\lambda$, it follows that all the coefficients $\xi_p$
vanish and that also $v=0$. This implies that
  \be  \sumpe p (\lAb{\omD 0}\pp,\Hb_\pp)
  = \sumqe{p'} (\lAbP{\omDP 0'}\qq,\HbP_\qq)  \,, \labl{2ss}
which is the desired generalization of \erf{neurel}. The sums in \erf{2ss}
are over the ideals of $\gb$ and $\gbP$, respectively.
{}From this result both the conjugacy class
selection rules and the invariance of the coset Virasoro algebra follow
in the same manner as in the case of diagonal cosets.

\Sect{Properties of the \smat\ of the coset theory}{scheck}

In this appendix we describe several checks which show that the \smat\
for the coset \cft, i.e.
  \be  \Scc\Lambda\LambdaP M{M'} =
  \Frac{|\GId|}{|\Gilabel|\cdot |\Gjlabel| } \sum_{\om\in\GId}
  \ch^*(\om)\, \CS^{[\om]}_{\ilabl,\jlabl} \, \cht(\om)  \labl{a1}
as defined in equation \erf{Scc} is a unitary symmetric matrix, that
its square gives a permutation of order 2 which leaves the vacuum fixed,
and that together with the $T$-matrix $\CT$ of \erf{tdef} it obeys
$(\CS\CT)^3 = \CS^2$, i.e.\ that it possesses all
 \futnot{the requirement that also the fusion coefficients are non-negative
integral is a requirement on the \cft, not on the \smat.}
the properties required for a proper \smat.

As a first step we derive some identities for the $S$-matrices of
\tbf s which will be helpful in proving these assertions.
Let us first derive some consequences of the unitarity of the tensor
product $S^\g_{}\otimes (S^{\g'}_{})^*_{}$
of the $S$-matrix of \g\ and the complex conjugate of the \smat\
of \gP.\,\futnote{Note, however, that this tensor product matrix does
{\em not\/} describe the behavior of the branching functions under
$\tau\mapsto-1/\tau$. This follows e.g.\ from the observation that
this matrix has non-vanishing matrix elements between vanishing and
non-vanishing branching functions.}
If $\ilabel$, $\klabel$ are the labels of two non-vanishing branching
functions,
then when summing over all elements of the identification group \Gid\ and over
all combinations $\jlabel$ of integrable weights of
$\g$ and $\gP$, the fact that $S^\g_{}\otimes (S^{\g'}_{})^*_{}$
is unitary implies that
  \be  \sum_{\omega\in\GId} \sum_{\jlabel}\!\! \CS_{\oMT\ilabel,\jlabel}
  \, \CS^*_{\jlabel,\klabel} = \sum_{\omega\in\GId}\!\!
  \delta_{\oMT\ilabel,\klabel }
  = |\Gilabel| \cdot \delta_{\ilabl,\klabl} \, , \labl{Gilabel}
where the last Kronecker delta refers to \Gid-orbits.
On the other hand, we can compute this sum also as
  \be \begin{array}{l}
  \dsum_{\omega\in\GId} \dsum_{\jlabel} \CS_{\omt\ilabel,\jlabel}\,
    \CS^*_{\jlabel,\klabel}  \\{}\\[-4mm] \hsp6
  = \dsum_{\omega\in\GId} \dsum_{\jlabel} \eE^{2\pi\ii Q_\omega(\jlabel)}\,
    \CS_{\ilabel,\jlabel}\, \CS^*_{\jlabel,\klabel} \\{}\\[-2mm] \hsp6
  = |\GId| \cdot
    \!\!\!\dsum_{\scriptstyle\jlabel \atop\scriptstyle Q(\jlabel)=0}\!\!\!
    \CS_{\ilabel,\jlabel}\, \CS^*_{\jlabel,\klabel}   \\{}\\[-3mm] \hsp6
  = \!\!\!\dsum_{\scriptstyle\jlabl \atop\scriptstyle Q(\jlabel)=0}\!\!\!
    \Frac{|\GId|^2}{|\Gjlabel|} \, \CS_{\ilabl,\jlabl}\, \CS^*_{\jlabl,\klabl}
  \,. \end{array} \ee
Here in the first line we used \erf{sjs}; then we performed the summation
over the identification group; and finally we wrote the result in terms of a
sum
over orbits of non-vanishing branching functions that have to be identified.
The notation $Q(\jlabl)=0$ indicates that the monodromy charge with respect to
all
simple currents in the identification group \Gid\ vanishes.
The \smat\ elements corresponding to {\em orbits\/} $\ilabl$ and $\jlabl$
of pairs $\ilabel$ and
$\jlabel$ appearing in the final form are well-defined because the original
\smat\ elements are constant on the orbits provided that their monodromy
charges vanish.

Comparison of the two results yields the identity
  \be  \!\!\dsum_{\scriptstyle\jlabl \atop\scriptstyle Q(\jlabel)=0}\!\!
    \Frac{|\GId|^2}{|\Gjlabel|} \cdot
    \CS_{\ilabl,\jlabl}\, \CS^*_{\jlabl,\klabl}
  = |\Gilabel| \cdot \delta_{\ilabl,\klabl} \, . \labl{usum}

Next we note that charge conjugation acts in a natural way on simple
current orbits, so that we are given a matrix $\CC_{\ilabl,\jlabl}$
which describes a permutation of order two of the orbits of
non-vanishing branching functions. Performing a similar
computation as above, one derives the identity
  \be  \!\!\dsum_{\scriptstyle\jlabl \atop\scriptstyle Q(\jlabel)=0}\!\!
    \Frac{|\GId|^2}{|\Gjlabel|} \cdot
    \CS_{\ilabl,\jlabl}\, \CS_{\jlabl,\klabl}
  = |\Gilabel| \cdot \CC_{\ilabl,\klabl} \, . \labl{csum}
analogous to \erf{usum}.

Furthermore, recall that we defined a $T$-matrix $\CT$ as the restriction to
orbits of non-vanishing branching functions of the product
$T^{\g}\otimes T^{\g'}$ of $T$-matrices of \g\ and \gP;
a similar computation as above leads to the relation
  \be  \!\!\dsum_{\scriptstyle\jlabl \atop\scriptstyle Q(\jlabel)=0}\!\!
    \Frac{|\GId|}{|\Gjlabel|} \,
    \CS_{\ilabl,\jlabl}^{}\, \CT_{\jlabl}^{}\, \CS_{\jlabl,\klabl}^{}
  = \CT_{\ilabl}^{-1}\, \CS_{\ilabl,\klabl}^{}\, \CT_{\klabl}^{-1} \,.
  \labl{msum}

Finally, in the derivation of the identities \erf{usum} -- \erf{msum} one
can substitute $\CS_{\ilabl,\jlabl}$ by $\CS^{[\omega]}_{\ilabl,\jlabl}$ and
$\CT_{\jlabl}$ by $\CT^{[\omega]}_{\jlabl}$, thereby obtaining
analogous relations for the matrices $\CS^{[\omega]}$ and $\CT^{[\omega]}$,
where the sum is now over all fields in the fixed point theory for which the
monodromy charge \wrt\ to all simple currents in the fixed point theory
vanishes,
$\QO(\ilabelO)=0$. However, as we have seen after equation \erf{qo2}, this
monodromy
charge vanishes precisely if the corresponding monodromy charge before
projection
vanishes. Thus the formulae \erf{usum} remains true if only $\CS$ is replaced
by
$\CS^{[\omega]}$. As for the analogue of equation \erf{msum} we remark that,
since the conformal weights of the fields into which a
fixed point is resolved differ only by integers, one has in fact
$\CT^{[\omega]}_{\jlabl}=\CT_{\jlabl}$. Also, one checks by inspection that
for all diagram automorphisms of \aff s
$\CC^{[\omega]}_{\ilabl,\klabl} = \CC_{\ilabl,\klabl}$ is
independent of $\omega$, too.
Therefore, we have formul\ae\ analogous to \erf{usum} -- \erf{msum} in which
only
$\CS$ is replaced by $\CS^{[\omega]}$.

We now combine the equations \erf{usum}, \erf{csum} and \erf{msum}
and their analogues for $\CS^{[\om]}$
instead of $\CS$ with the following two identities for the characters of
an abelian group. First,
  \be \sum_{\ch\in\Gilabels}\!\! \ch(\omega)= |\Gilabel|\cdot
  \delta_{\omega, e} \,, \ee
where $e$ denotes the unit element of the group $\Gilabel \subseteq \Gid$.
Second, for any $\ch\iN\Gilabels$ and any $\cht\iN\Gjlabels$ one has
  \be  \sum_{\omega\in\GId} \ch^*(\omega)\,\cht(\omega) =
  \sum_{\omega\in \Gilabel\cap\Gjlabel} \ch^*(\omega)\,\cht(\omega) =
  |\Gilabel\,\cap\Gjlabel|\cdot \delta_{\ch,\cht} \, .  \ee

To show that the \smat\ is symmetric, we take the transpose of \erf{a1}.
Observing that $\omD$ and $\omD^{-1}$ yield the same \olie\ so that
  \be  \CS^{[\omega^{-1}]}_{\ilabel,\jlabel}
  =  \CS^{[\omega]}_{\ilabel,\jlabel} \, , \ee
and that this is a symmetric matrix, we get
  \be  \Scc M{M'}\Lambda\LambdaP =
  \Frac{|\GId|}{|\Gilabel|\cdot |\Gjlabel| } \sum_{\om\in\GId}
  \cht^*(\om)\, \CS^{[\om^{-1}]}_{\ilabl,\jlabl} \, \ch(\om)  \,. \ee
Using the property $\ch(\omega^{-1}) = \ch^*(\omega)$
of group characters, the symmetry property then follows by summing over
$\omega^{-1}$ instead of over $\omega$.

To show that the \smat\ \erf{Scc} is unitary, we calculate
  \be\begin{array}l
  \dsum_{\jlabl}\dsum_{\chz\in\Gjlabels} \CS_{(\ilabl,\che),(\jlabl,\chz)}
  (\CS_{(\jlabl,\chz),(\klabl,\chd)})^* \\[1mm]\hsp4
  = \dsum_{\jlabl}
  \Frac{|\GId|^2}{|\Gilabel|\cdot |\Gjlabel|^2\cdot |\Gklabel|}
  \dsum_{\chz\in\Gjlabels} \dsum_{\ome,\omz\in\GId}
  \CS^{[\ome]}_{\ilabl, \jlabl} (\CS^{[\omz]}_{\jlabl,\klabl})^* \\[.3em]
  \hsp{24.2} \times \che^*(\ome)\chz(\ome)\chz(\omz)\chd^*(\omz) \\[1mm]\hsp4
  = \dsum_{\jlabl}
  \Frac{|\GId|^2}{|\Gilabel|\cdot |\Gjlabel|\cdot |\Gklabel|}
  \dsum_{\ome\in\GId}
  \CS^{[\ome]}_{\ilabl, \jlabl} (\CS^{[\ome]}_{\jlabl,\klabl})^*\,
  \che^*(\ome) \chd^*(\ome^{-1}) \\{}\\[-3mm]\hsp4
  = \Frac1{|\Gklabel|} \dsum_{\ome\in\GId} \delta_{\ilabl,\klabl}
  \che^*(\ome) \chd(\ome) \\{}\\[-3mm]\hsp4
  = \delta_{\ilabl,\klabl}\, \delta_{\che,\chd} \, . \end{array}\ee
Here we used the analogue of the relation \erf{usum};
a parallel computation using \erf{csum} shows that
  \be \sum_{\jlabl}\sum_{\chz\in\Gjlabels} \CS_{(\ilabl,\che),(\jlabl,\chz)}
  \CS_{(\jlabl,\chz),(\klabl,\chd)} = \CC_{[\ilabel],[\klabel]}\,
  \delta_{\che,\chd} \, .\ee
Here $\CC_{[\ilabel],[\klabel]}$ is the charge conjugation matrix of the
coset theory; the matrix $\CC$ is manifestly a permutation of order two of
the primary fields which leaves the identity primary field $[0;0]$
fixed. Thus charge conjugation acts in a
trivial way on the labels which describe the fixed point resolution.

Our final check concerns the relation $(\CS\CT)^3=\CS^2$, or equivalently,
$\CS\CT\CS=\CT^{-1} \CS \CT^{-1}$. Using \erf{msum} we have
  \be \begin{array}l
  \dsum_{\jlabl}\dsum_{\chz\in\Gjlabels} \CS_{(\ilabl,\che),(\jlabl,\chz)}
  \CT_{\jlabl} \CS_{(\jlabl,\chz),(\klabl,\chd)} \\{}\\[-3mm]\hsp4
  = \dsum_{\jlabl}
  \Frac{|\GId|^2}{|\Gilabel|\cdot |\Gjlabel|^2\cdot |\Gklabel|}
  \dsum_{\chz\in\Gjlabels}
  \\ \hsp9 \times
  \dsum_{\ome,\omz\in\GId}
  \CS^{[\ome]}_{\ilabl,\jlabl} \CT_{\jlabl}^{} \CS^{[\omz]}_{\jlabl,\klabl}\,
  \che^*(\ome)\chz(\ome)\chz^*(\omz)\chd(\omz) \\{}\\[-3mm]\hsp4
  = \dsum_{\jlabl}
  \Frac{|\GId|^2}{|\Gilabel|\cdot |\Gjlabel|\cdot |\Gklabel|} \cdot
  \dsum_{\ome\in\GId} \CS^{[\ome]}_{\ilabl,\jlabl}\CT_{\jlabl}^{}
  \CS^{[\ome]}_{\jlabl,\klabl}\, \che^*(\ome) \chd(\ome) \\{}\\[-3mm]\hsp4
  = \Frac{|\GId|}{|\Gilabel|\cdot |\Gklabel|} \dsum_{\ome\in\GId}
  \CT_{\ilabel}^{-1} \CS^{[\ome]}_{\ilabel, \klabel} \CT_{\klabel}^{-1}
  \che^*(\ome) \chd(\ome) \\{}\\[-3mm]\hsp4
  = \CT_{\ilabl}^{-1}\, \CS_{(\ilabl,\che),(\klabl,\chd)}\,\CT_{\klabl}^{-1}\,.
  \end{array}\ee

In summary, for any coset \cft\ in which the identification group $\Gid$
possesses the \facprop\ we obtain a unitary \rep\ of \slz, with
$\CS$ symmetric and $\CT$ diagonal.

Let us also mention that we do not yet have a general proof
that inserting the formula \erf{a1} into the Verlinde formula yields
non-negative integral fusion coefficients. However,
we have checked in many examples that this further condition for
the consistency of the coset \cft\ is indeed fulfilled.

\vskip3em

  \newcommand{\wb}{\,\linebreak[0]} \def\wB {$\,$\wb}
  \newcommand{\Bi}[1]    {\bibitem{#1}}
  \newcommand{\Erra}[3]  {\,[{\em ibid.}\ {#1} ({#2}) {#3}, {\em Erratum}]}
  \newcommand{\BOOK}[4]  {{\em #1\/} ({#2}, {#3} {#4})}
  \newcommand{\vypf}[5]  {\ {\sl #5}, {#1} [FS{#2}] ({#3}) {#4}}
  \renewcommand{\J}[5]     {\ {\sl #5}, {#1} {#2} ({#3}) {#4} }
  \newcommand{\Prep}[2]  {{\sl #2}, preprint {#1}}
\def\jf            {J.\ Fuchs}
 \def\adma  {Adv.\wb Math.}
 \def\anop  {Ann.\wb Phys.}
 \def\foph  {Fortschr.\wb Phys.}
 \def\hepa  {Helv.\wb Phys.\wB Acta}
 \def\ijmp  {Int.\wb J.\wb Mod.\wb Phys.\ A}
 \def\jopa  {J.\wb Phys.\ A}
 \def\mpla  {Mod.\wb Phys.\wb Lett.\ A}
 \def\nn    {$N=2$ }
 \def\npbF  {Nucl.\wb Phys.\ B\vypf}
 \def\npbp  {Nucl.\wb Phys.\ B (Proc.\wb Suppl.)}
 \def\nuci  {Nuovo\wB Cim.}
 \def\nupb  {Nucl.\wb Phys.\ B}
 \def\phlb  {Phys.\wb Lett.\ B}
 \def\phrd  {Phys.\wb Rev.\ D}
 \def\prep  {Phys.\wb Rep.}
 \def\comp  {Com\-mun.\wb Math.\wb Phys.}
 \def\jomp  {J.\wb Math.\wb Phys.}

 \def\A       {Algebra}
 \def\alg     {algebra}
 \def\Be     {{Berlin}}
 \def\BIR    {{Birk\-h\"au\-ser}}
 \def\Ca     {{Cambridge}}
 \def\class   {classification }
 \def\compac  {compactification}
 \def\con     {conformal\ }
 \def\CUP    {{Cambridge University Press}}
 \def\furu    {fusion rule}
 \def\GB     {{Gordon and Breach}}
 \def\ide     {identification}
 \newcommand{\inBO}[7]  {in:\ {\em #1}, {#2}\ ({#3}, {#4} {#5}),  p.\ {#6}}
 \def\Infdim  {Infinite-dimensional}
 \def\inv     {invariance}
 \def\modinvt  {modular invariant}
 \def\NY     {{New York}}
 \def\parfu   {partition function}
 \def\Q       {Quantum\ }
 \def\qg      {quantum group}
 \def\Rep     {Representation}
 \def\SV     {{Sprin\-ger Verlag}}
 \def\syms    {sym\-me\-tries}
 \def\wzw     {WZW\ }
 \def\va      {Virasoro algebra}

\version\versionno \end{document}